\documentclass[11pt,a4paper]{article}

\newtheorem{lemma}{Lemma}
\newtheorem{theorem}{Theorem}
\newtheorem{corollary}{Corollary}

\def \QED{\mbox{\rule[0pt]{0.7em}{0.7em}}}

\usepackage{cite}
\usepackage{easybmat}
\usepackage{amsmath}
\usepackage{amssymb}
\usepackage{graphicx}

\title{ \bf Distributed Blind Calibration via Output Synchronization in Lossy Sensor Networks}

\author{Milo\v{s} S. Stankovi\'{c}\thanks{Innovation Center, School of Electrical Engineering, University of Belgrade,
Serbia (e-mail: milos.stankovic@ic.etf.rs).}, \hspace{2mm}Srdjan  S. Stankovi\'{c}\thanks{School of Electrical Engineering, University of Belgrade, Serbia (e-mail: stankovic@etf.rs).}\hspace{2mm} and \\ Karl Henrik Johansson\thanks{ACCESS Linnaeus Center, School of Electrical Engineering, KTH Royal Institute of Technology, Stockholm, Sweden (e-mail: kallej@kth.se).} }

\begin{document}

\maketitle
\let\thefootnote\relax\footnote{Preliminary results of this work have been presented at the IEEE Conf. Decision and Control 2012 \cite{calcdc12} and the Mediterranean Conference on Control and Automation 2012 \cite{calmed}.}
\let\thefootnote\relax\footnote{This work was supported by the EU Marie Curie CIG, Knut and Alice Wallenberg
Foundation and the Swedish Research Council.}

\begin{abstract} In this paper a novel distributed algorithm for blind macro calibration in sensor networks based on output synchronization is proposed. The
algorithm is formulated as a set of gradient-type recursions for estimating parameters of
sensor calibration functions, starting from local criteria defined as weighted sums of mean square differences between the outputs of
neighboring sensors. It is
proved, on the basis of an originally developed methodology for treating higher-order consensus (or output synchronization) schemes, that the algorithm achieves asymptotic agreement for sensor gains and offsets, in the mean square sense and with probability one. In the case of additive measurement noise, additive inter-agent communication noise, and communication outages, a modification  of the original algorithm based on instrumental variables is proposed. It is
proved using stochastic approximation arguments that the modified algorithm achieves asymptotic consensus for sensor gains and offsets, in
the mean square sense and with probability one. Special attention is paid to the situation when a subset of sensors in the network remains with fixed characteristics. Illustrative simulation examples are provided.

\end{abstract}


\section{Introduction}

Recently, \emph{wireless sensor networks} (WSN's) have emerged as an important research area (see, \emph{e.g.}, \cite{wsn,assc,sfj}). Diverse new applications have sparked the recognition of new classes of problems for the developers and users. \emph{Sensor calibration} represents one of the most important challenges for the wide deployment of this new technology. Only relatively small sensor-systems can utilize micro-calibration, in which each device is individually tuned in a controlled environment. Larger WSN's demand new concepts and methods for calibration, since many
devices can be in partially unobservable and dynamic environments. The so-called \emph{macro-calibration} is based on the idea to
calibrate a \emph{network as a whole} by observing only the overall system response, thus eliminating the need
to directly calibrate each and every device. The usual prerequisite is to frame calibration as a parameter
estimation problem (\textit{e.g.}, \cite{whcu1,whcu2}). Automatic methods for jointly calibrating WSN's, without dependence on
controlled stimuli or high-fidelity groundtruth data, are of significant interest. The underlying practical idea is to have sensors with homogeneous properties irrespective of the lack of an absolute reference, expecting (or imposing) dominant influence of well calibrated sensors. This problem is referred to as
\emph{blind calibration}. In \cite{balnow1,balnow2} a \emph{centralized}, non-recursive algorithm was proposed for blind calibration assuming non-identical but correlated sensor readings. Another approach to blind WSN calibration is to assume that the deployment is relatively dense, so that neighboring nodes have nearly identical readings. In \cite{bmep} this scenario was adopted, but the algorithm assumes only \emph{pairwise} inter-node communications. There are also methods trying to cope with situations in which sensor network deployments may not meet the density requirements as, \textit{e.g.}, \cite{tcs} where the calibration is done by sudden measurement error correction based on training and running algorithm phases.

In this paper we propose a novel \emph{blind macro-calibration method} for sensor networks based on \emph{distributed on-line estimation}
of the parameters of affine \emph{calibration functions}. It is assumed that the sensors
form a network with \emph{directed communication links} between neighboring nodes. By formulating the calibration problem as the problem of distributed minimization of a set of weighted sums of mean square differences between the outputs of neighboring sensors, we derive a \emph{distributed gradient type recursive algorithm} and show that the overall network
behavior can be treated as a nontrivial \textit{output synchronization} or \emph{consensus} problem, in which all the equivalent sensor gains and
offsets should \emph{converge asymptotically to equal values}. However, formally speaking, properties of the method cannot be directly analyzed using the classical results related to different versions of the
dynamic consensus and synchronization algorithms (see, \textit{e.g.}, \cite{osfm,rb, Wieland20111068,6426752} with the references therein, together with a discussion given below in Section 4). To the authors' best knowledge, consensus has been applied directly to the
calibration problems only in \cite{bolognani2009,calibree}, but within different contexts. Using general arguments
related to stability of diagonally dominant dynamic systems \cite{ohtasiljak,s,pierce}, a basic lemma is derived dealing with characteristics of matrices decomposed into blocks, applicable to a large class of higher order consensus schemes. Using this result, we prove that the
proposed basic algorithm achieves \emph{asymptotic consensus for sensor gains and offsets} in the mean square sense and with probability one (w.p.1). The asymptotic gains and offsets depend, in general, on initial conditions, together with signal, sensor and network characteristics; the freedom of choosing the weights in the local criterion functions (which define the level of confidence of particular sensors) offers a great flexibility to the designer and a possibility to obtain good performance in a large variety of situations in practice. The basic results are then extended by assuming: 1) additive  communication noise, 2) communication outages, and 3) additive measurement noise. An algorithm of
\emph{instrumental variable} type \cite{sods} is proposed for solving the problem posed in this case, and the
achievement of the asymptotic consensus in the mean square sense and w.p.1 is proved for both gains and offsets. In case when a subset of nodes in the network is kept with fixed (acceptable) characteristics, it is proved that the algorithm provides convergence in the mean square sense and w.p.1 to specific points representing ``centroids'' of these characteristics, defined by the the selected weights and network properties. It is to be emphasized that the proposed methodology does not require any supervision or fusion center.

The outline of the paper is as follows. The following subsection introduces notation and basic definitions to be used throughout the paper. In Section 2 we formulate the blind calibration problem and introduce the basic algorithm. Section 3 is devoted to the algorithm's convergence analysis in the noiseless case, under different assumptions on the measured signals and network structure. In Section 4 a modified algorithm with instrumental variables is proposed for the general case of lossy sensor networks. Its convergence in the mean square sense and w.p.1 is proved and the convergence rate estimated. Section 5 is devoted to the convergence analysis of the proposed algorithm in the case of macro-calibration where a subset of sensors remains with fixed characteristics. In Section 6 we present some illustrative simulation results.

\subsection{Notation and some Definitions}

$\mathbb{R}$ denotes the set of real numbers, $\mathbb{C}$ denotes the set of complex numbers, while $\mathbb{C}_+$ denotes the set of complex numbers with positive real parts. $E\{\cdot\}$ denotes the mathematical expectation. $I_k$ denotes the identity matrix of dimension $k$, $0_{i\times j}$ denotes a $i \times j$ matrix with all the elements equal to zero. $\otimes$ denotes the Kronecker product. ${\rm diag}\{\ldots\}$ denotes a block diagonal matrix with the specified diagonal elements. $\|\cdot\|$ denotes an operator norm. ${\rm span}\{A\}$ denotes a linear space spanned by the columns of matrix $A$. $\lambda_{\min}(A)$ denotes an eigenvalue of matrix $A$ with the smallest absolute value, while $\lambda_{\max}(A)$ denotes an eigenvalue of matrix $A$ with the largest absolute value.
\par
Matrix $A \in \mathbb{R}^{k\times k}$ is said to be an M-matrix if its off-diagonal entries are less than or equal to zero, and if all its principal minors are positive (see, \textit{e.g.}, \cite{s} for different equivalent characterizations of M-matrices).
\par
In a given directed graph (digraph) $\mathcal{G}=(\mathcal{N},\mathcal{E})$, where $\mathcal{N}$ is the set of nodes (vertices) and $\mathcal{E}$ is the set of links (arcs), if there is a walk from the node $j$ to the node $i$ we say that the node $i$ is reachable from $j$. A node from which every node in the digraph is reachable is called center node.

\section{Problem Formulation and the Main Algorithm}
Consider $n$ distributed sensors measuring a discrete-time signal $x(t)$,  $t=\ldots, -1,0, 1, \ldots$,
and assume that the output of the $i$-th sensor can be represented by
\begin{equation} \label{yi}
y_{i}(t)=\alpha_{i} x(t) + \beta_{i},
\end{equation}
where the gain $\alpha_{i}$ and the offset $\beta_{i}$ are unknown constants.
\par By sensor calibration we consider the application of an affine \emph{calibration
function} which produces the overall sensor output
\begin{equation} \label{calfun}
z_{i}(t)=a_{i}y_{i}(t)+b_{i} =  g_{i} x(t)+ f_{i},
\end{equation}
where $a_{i}$ and $b_{i}$ are the calibration parameters, $g_{i}=a_{i} \alpha_{i}$ is the equivalent  gain and $f_{i}=a_{i} \beta_{i}+ b_{i}$ the equivalent offset. In general, parameters $a_{i}$ and $b_{i}$  have to be chosen in such a way as to set
$g_{i}$ as close as possible to one and $f_{i}$ as close as possible to zero.
\par
We assume that the observed sensors form a network with a predefined structure of inter-sensor
communications represented by a \emph{directed graph}
$\mathcal{G}=(\mathcal{N},\mathcal{E})$, where $\mathcal{N}$ is the set of nodes (one node corresponds to one
sensor) and $\mathcal{E}$ the set of links (arcs). Define the adjacency matrix
$A=[a_{ij}]$, $i,j=1, \ldots, n$, where $a_{ij}=1$ if the $j$-th sensor can send its message to
the $i$-th sensor, and $a_{ij}=0$ otherwise. Let $\mathcal{N}_{i}$ be the
set of neighboring nodes of the $i$-th node, \emph{i.e.}, the set of nodes $j$ for which $a_{ij}=1$.

Starting from the general concept of \emph{blind macro calibration}, the aim of this paper is to propose an algorithm for \emph{distributed real-time estimation of the calibration
parameters} $a_{i}$ and $b_{i}$  which would provide asymptotically equal outputs $z_{i}(t)$ of all the sensors without the knowledge of the measured signal. Furthermore, the algorithm should be adjusted such that, loosely speaking, the well calibrated sensors would correct, on the basis of global \emph{consensus}, the behavior of those that are not.
Assuming that $\{x(t) \}$ is a stochastic process,
the algorithm will be derived starting from the following set of \emph{local
criteria}:
\begin{align} \label{J}
J_{i}= \sum_{j \in \mathcal{N}_{i}}  \gamma_{ij} E \{(z_{j}(t)-z_{i}(t))^{2}\},
\end{align}
$i=1,..., n$, where
$\gamma_{ij}$ are nonnegative scalar weights reflecting, in general,
relative importance of the neighboring nodes. Denoting $\theta_{i}=[a_{i} \;\; b_{i}]^{T}$, we obtain for the gradient of \eqref{J}:
\begin{equation} \label{grad}
{\rm grad}_{\theta_{i}} J_{i}=\sum_{j \in \mathcal{N}_{i}}  \gamma_{ij}E\{(z_{j}(t)-z_{i}(t))\left[
\begin{BMAT}{c}{cc} y_{i}(t) \\ 1 \end{BMAT} \right]\}.
\end{equation}
The last equation gives rise to the following
stochastic \textit{gradient recursion} for estimating $\theta_{i}^{*}$ which minimizes (\ref{J})
\begin{equation} \label{alg}
\hat{\theta}_{i}(t+1)=\hat{\theta}_{i}(t)+ \delta_{i}(t) \sum_{j \in \mathcal{N}_{i}}\gamma_{ij}
\epsilon_{ij}(t) \left[
\begin{BMAT}{c}{cc} y_{i}(t) \\ 1 \end{BMAT} \right],
\end{equation}
where  $\hat{\theta}_{i}(t)=[\hat{a}_{i}(t) \;\; \hat{b}_{i}(t)]^{T}$, $\delta_{i}(t) > 0$ is a time-varying
gain influencing convergence properties of the algorithm, $\epsilon_{ij}(t)=\hat{z}_{j}(t)-\hat{z}_{i}(t)$ and
$\hat{z}_{i}(t)=\hat{a}_{i}(t) y_{i}(t)+\hat{b}_{i}(t)$, with the initial conditions $\hat{\theta}_{i}(0)=[1 \;
\; 0]^{T}$, $i=1, \ldots, n$. Notice that each iteration of the algorithm (\ref{alg}) subsumes availability of the local measurement and the reception of current
messages of the neighboring nodes' outputs $z_j(t)$ (communication outages will be discussed later). Local computational efforts for individual agents are minor, having in mind that only two parameters are estimated. Communication efforts 
depend on the number of agents in the neighborhoods, which is kept to be small in actual WSN's.  

The underlying idea of the set of recursions (\ref{alg}) is to ensure that the estimates of all the local
gains $\hat{g}_{i}(t)=\hat{a}_{i}(t) \alpha_{i}$ and offsets $\hat{f}_{i}(t)=\hat{a}_{i}(t)
\beta_{i}+\hat{b}_{i}(t)$ tend asymptotically to the \emph{same values} $\bar{g}$ and $\bar{f}$, respectively, implying
$\hat{z}_{j}(t)=\hat{z}_{i}(t)$, $i,j=1, \ldots, n$. For the sake of compact notations, introduce
\begin{equation} \label{rho_i}
\hat{\rho}_{i}(t) =\left[
\begin{BMAT}{c}{cc} \hat{g}_{i}(t) \\ \hat{f}_{i}(t) \end{BMAT} \right]=\left[
\begin{BMAT}{cc}{cc} \alpha_{i}  &  0  \\\beta_{i}& 1 \end{BMAT} \right] \hat{\theta}_{i}(t),
\end{equation}
and
\begin{equation}
\epsilon_{ij}(t)=\left[
\begin{BMAT}{cc}{c} x(t) & 1 \end{BMAT} \right] (\hat{\rho}_{j}(t)-\hat{\rho}_{i}(t)),
\end{equation}
so that (\ref{alg}) becomes
\begin{equation} \label{algro}
\hat{\rho}_{i}(t+1)=\hat{\rho}_{i}(t)+ \delta_{i}(t) \sum_{j \in \mathcal{N}_{i}}\gamma_{ij} \Phi_{i}(t)
(\hat{\rho}_{j}(t)-\hat{\rho}_{i}(t)),
\end{equation}
where
\begin{align} \Phi_{i}(t)=& \left[
\begin{BMAT}{cc}{cc} \alpha_{i}y_{i}(t)x(t)  &  \alpha_{i} y_{i}(t) \\  [1+\beta_{i}y_{i}(t)]x(t) & 1 + \beta_{i} y_{i}(t) \end{BMAT}
\right]  \\=& \left[
\begin{BMAT}{cc}{cc} \alpha_{i}\beta_{i}x(t)+ \alpha_{i}^{2}x(t)^2  &  \alpha_{i}\beta_{i}+  \alpha_{i}^2 x(t) \\
 (1+\beta_{i}^2)x(t) + \alpha_{i}\beta_{i}x(t)^2 & 1 + \beta_{i}^2+\alpha_{i}\beta_{i}x(t) \end{BMAT}
\right], \nonumber
\end{align}
with the initial conditions $\hat{\rho}_{i}(0)=[\alpha_{i} \; \; \beta_{i}]^{T},$  $i=1, \ldots, n$. Recursions (\ref{algro}) can be represented in the following compact form
\begin{equation} \label{compalg}
\hat{\rho}(t+1)=[I+ (\Delta(t) \otimes I_{2}) B(t)] \hat{\rho}(t),
\end{equation}
where $\hat{\rho}(t)= [\hat{\rho}_{1}(t)^{T} \cdots \hat{\rho}_{n}(t)^{T} ]^{T},$ $\Delta(t)= {\rm diag} \{
\delta_{1}(t),$ $ \ldots, \delta_{n}(t) \}$, \[B(t)= \Phi(t)(\Gamma \otimes I_{2}), \]
   $\Phi(t)= {\rm diag} \{\Phi_{1}(t),\ldots, \Phi_{n}(t) \}$,
\[\Gamma =\left[
\begin{BMAT}{cccc}{cccc} -\sum_{j, j \neq 1} \gamma_{1j}  &  \gamma_{12} & \cdots & \gamma_{1n} \\ \gamma_{21} & -\sum_{j, j \neq 2}
\gamma_{2j} & \cdots & \gamma_{2n} \\ & & \ddots & \\ \gamma_{n1} & \gamma_{n2} & \cdots & -\sum_{j, j \neq n}
\gamma_{nj}  \end{BMAT} \right],\] where $\gamma_{ij}=0$ when $j \notin \mathcal{N}_{i}$; the initial
condition is $\hat{\rho}(0)= [\hat{\rho}_{1}(0)^{T} \cdots \hat{\rho}_{n}(0)^{T} ]^{T}, $ in accordance with
(\ref{algro}). The asymptotic value of $\hat{\rho}(t)$, which depends on the initial conditions and
matrix $B(t)$, which is, in turn, a function of the signal, sensor and network parameters, should be such that the
components with odd indices and the components with even indices become equal.
\par
Notice that, in general, the choice of the weighting coefficients $\gamma_{ij}$ plays an important role in achieving successful performance of the whole scheme. If the underlying idea of the whole methodology is to achieve good absolute calibration results ($\bar{g}$ as close as possible to one and $\bar{f}$ as close as possible to zero) by exploiting sensors with good characteristics in a large sensor network, there are two main possibilities: 1)  to rely on the majority of good sensors, when all $\gamma_{ij}$ in any neighborhood $\mathcal{N}_{i}$ can take the same value, or 2) to emphasize the effect of \emph{a priori} selected good sensors $j$ belonging to a set $  \mathcal{N}^{f} \subset \mathcal{N} $ by: a) attaching to them relatively high values of $\gamma_{ij}$ within all $\mathcal{N}_{i}$ for which $i$ is an out-neighbor of $j$, b) multiplying all $\gamma_{jk}$, $k=1,...,n$, $k \neq j$, by a relatively small positive number (thus preventing big changes of $\hat{\rho}_j(t)$). Section 5 will be devoted to the situation in which a set of reliable sensors in the network is left with fixed characteristics. 

\section{Convergence Analysis - Noiseless Case}

In this section  we assume no communication errors and no measurement errors in order to emphasize structural properties of the proposed algorithm. For this scenario constant step size in the derived recursions is assumed:
\par
A1) $\delta_{i}(t)=\delta= {\rm const}$, $i=1, \ldots, n$.

For the sake of clearer presentation, we first adopt a simplifying assumption, which will be relaxed later:

A2) $\{ x(t) \}$ is i.i.d., with $E \{ x(t) \}= \bar{x} < \infty$ and $E \{ x(t)^2 \}=s^2 < \infty$.

Based on A1) and A2), we obtain the following recursion for the mean of the parameter estimates $ \bar{\rho}(t)=E \{ \rho(t) \}$

\begin{equation} \label{compalgm}
\bar{\rho}(t+1)=(I+ \delta \bar{B}) \bar{\rho}(t),
\end{equation}
where  $ \bar{\rho}(0)= \rho(0) $,  $\bar{B}= \bar{\Phi}(\Gamma \otimes I_{2})$
and $\bar{\Phi}= E \{ \Phi(t) \}={\rm diag} \{ \bar{\Phi}_{1} \ldots \bar{\Phi}_{n} \} $, with
\[ \bar{\Phi}_{i}=\left[ \begin{BMAT}{cc}{cc} \alpha_{i} \beta_{i}\bar{x}+ \alpha_{i}^{2} s^{2} &\alpha_{i} \beta_{i} + \alpha_{i}^{2} \bar{x}
 \\ (1+\beta_{i}^{2}) \bar{x} + \alpha_{i} \beta_{i} s^{2} &
 1+ \beta_{i}^{2}+\alpha_{i} \beta_{i} \bar{x}\end{BMAT}
\right]. \]

A closer insight into the recursion (\ref{compalgm}) shows that its properties can be analyzed neither by applying the well known
results related to the classical first-order consensus schemes, nor those related to special forms of second-order consensus schemes, due to the block structure of $\bar{B}$ composed of specific $2 \times 2$ block matrices (see, \textit{e.g.}, \cite{osfm,rb, liuanderson, zanella} and many references therein). In order to cope with this problem, we shall first propose a novel methodology based on the concept of
 \emph{diagonal dominance of matrices decomposed into blocks} \cite{ohtasiljak,s}. We formulate several lemmas which represent basic prerequisites for the subsequent analysis of the proposed algorithm.

\begin{lemma} \label{lemma:1} \cite{ohtasiljak,pierce} A matrix $A = [A_{ij}]$, where $A_{ij} \in \mathbb{C}^{m \times m }$,
$i,j=1, \ldots n$, has \emph{quasi-dominating diagonal blocks} if the test matrix $W \in \mathbb{R}^{n \times
n }$, with the elements
\[w_{ij}=1 \;\; (i=j); \;\;\;  w_{ij}=-\| A_{ii}^{-1} A_{ij} \| \;\; (i \neq j) \] is an M-matrix; then, $A$ is nonsingular. If $A- \lambda I$ has quasi-dominating diagonal blocks for all $\lambda \in
\mathbb{C}_{+} $, then $A$ is Hurwitz. \end{lemma}
\par
The following lemma has an important role, and represents \emph{per se} one of the contributions of this paper.

\begin{lemma} \label{lemma:2} If $A = [A_{ij}]$ has quasi-dominating diagonal blocks and $A_{ii}$,  $i=1, \ldots, n$, are Hurwitz, $A$ is
also Hurwitz. \end{lemma}

The proof, based on the basic ideas exposed in \cite{ohtasiljak,pierce}, is given in the Appendix. As will be seen below, the general result of Lemma~\ref{lemma:2} will be utilized only in the context of the proposed algorithm, in relation with $2 \times 2$ blocks $A_{ij}$. It is hope that it can be found useful in analyzing stability of higher order dynamic consensus schemes.

The structure of the blocks of matrix $\bar{B}$ is defined by the structure of the digraph $\mathcal{G}$. We assume the following assumption, typical for consensus schemes:

A3) Graph $\mathcal{G}$ has a spanning tree.

Assumption A3) implies that matrix $\Gamma$ has one eigenvalue at the origin and the other eigenvalues with
negative real parts, \emph{e.g.}, \cite{rb}. Consequently, matrix $\bar{B}$ from (\ref{compalgm}) has at least two eigenvalues at the origin. In the following, we characterize its remaining eigenvalues.
\par
A4) $s^{2} - \bar{x}^{2}= {\rm var} \{ x(t) \}
 > 0$.
\par
Assumption A4) essentially ensures sufficient excitation by the measured signal, which should not remain constant. Its important direct formal consequence is that $-\bar{\Phi}_{i}$ is Hurwitz, $i=1, \ldots, n$. Namely, it is straightforward to verify that $-\bar{\Phi}_{i}$ is Hurwitz iff
\begin{equation}
\alpha_{i}^{2}(s^{2}-\bar{x}^{2}) > 0, \;\;\; 2\alpha_{i} \beta_{i} \bar{x} +\alpha_{i}^{2}s^{2}+1+\beta_{i}^{2} > 0.
\end{equation}
Both inequalities hold iff A4) holds.
\par
The following lemma applies the above general results to matrix $\bar{B}$; it is essential for all the derivations which follow.
\par
\begin{lemma} \label{lemma:3} Let Assumptions A3) and A4) be satisfied.
Then, matrix $\bar{B}$ in (\ref{compalgm}) has two eigenvalues at the origin and the remaining eigenvalues have
negative real parts. \end{lemma}
\par
The proof is given in the Appendix.
\par
Define vectors $i_{1}= \left[ \begin{BMAT}{ccccccc}{c} 1 & 0 & 1 & 0 & \ldots & 1 & 0 \end{BMAT} \right]^{T} \in \mathbb{R}^{2n}$ and
$i_{2}= \left[
\begin{BMAT}{ccccccc}{c} 0 & 1 & 0 & 1 & \ldots & 0 & 1 \end{BMAT} \right]^{T} $ $ \in \mathbb{R}^{2n}$ which represent the right eigenvectors of $\bar{B}$ corresponding to the zero eigenvalue, and let $\pi_{1}$ and $\pi_{2}$ be the corresponding normalized left eigenvectors, satisfying $\left[\begin{BMAT}{c}{c.c} \pi_{1} \\ \pi_{2}   \end{BMAT}  \right] \left[\begin{BMAT}{c.c}{c} i_{1} & i_{2}   \end{BMAT} \right]=I_2$. The following lemma introduces a similarity transformation which will be used in the rest of the derivations.
\par
\begin{lemma} \label{lemma:4} Let $T= \left[
\begin{BMAT}{c.c.c}{c} i_{1} & i_{2} & T_{2n \times (2n-2)} \end{BMAT} \right]$, where $T_{2n \times (2n-2)}$ is an $2n \times (2n-2)$ matrix, such that span$\{T_{2n \times (2n-2) }\}
$= span$\{\bar{B}\}$. Then, $T$ is nonsingular and
\begin{equation} \label{tminusone}
T^{-1} \bar{B} T= \left[ \begin{BMAT}{c.c}{c.c} 0_{2 \times 2} & 0_{2 \times (2n-2)}   \\ 0_{(2n-2) \times 2} &
\bar{B}^{*}
\end{BMAT} \right],
\end{equation}
where $\bar{B}^{*}$ is Hurwitz.
\end{lemma}
\par
The proof is given in the Appendix. Notice that
\begin{equation} \label{trans}
T^{-1}=\left[ \begin{BMAT}{c}{c.c.c} \pi_{1} \\ \pi_{2} \\ S_{(2n-2) \times 2n}
\end{BMAT} \right],
\end{equation}
where $S_{(2n-2) \times 2n }$ is defined in accordance with the definition of $T$.

We are now ready to prove the following theorem dealing with the asymptotic behavior of the mean of the parameter estimates generated by \eqref{compalgm}.

\begin{theorem} \label{theorem:1} Let Assumptions A1), A2), A3) and A4) be satisfied. Then there exists $\delta'
> 0$ such that for all $\delta \leq \delta'$ in (\ref{compalgm}), $\bar{\rho}_{\infty}=\lim_{t \to \infty} \bar{\rho}(t)=(i_{1} \pi_{1}+i_{2} \pi_{2}) \bar{\rho}(0)$, implying $\bar{\rho}_{\infty}=[\bar{\rho}_{\infty 1}^{T} \cdots \bar{\rho}_{\infty n}^{T} ]^{T}$ with $\bar{\rho}_{\infty 1 }=\bar{\rho}_{\infty 2}= \cdots =
\bar{\rho}_{\infty n }$.
\end{theorem} {\bf Proof:} Let $\tilde{\bar{\rho}}(t)=$$  [\tilde{\bar{\rho}}_{1}(t)^{T}  \cdots
\tilde{\bar{\rho}}_{n}(t)^{T} ]^{T}$=$T^{-1} \bar{\rho}(t)$. From (\ref{compalgm}) we obtain
\begin{equation} \label{onetwo}
 \tilde{\bar{\rho}}(t+1)^{[1]}=\tilde{\bar{\rho}}(t)^{[1]};  \;\;\;
 \tilde{\bar{\rho}}(t+1)^{[2]}=(I+\delta \bar{B}^{*})  \tilde{\bar{\rho}}(t)^{[2]},
\end{equation}
where ${\rm dim} \{ \tilde{\bar{\rho}}(t)^{[1]}\}=2$, $
{\rm dim}\{ \tilde{\bar{\rho}}(t)^{[2]}\}=2n-2$. Using Lemma~\ref{lemma:3} we conclude that for $\delta$
small enough all the eigenvalues of $I+\delta \bar{B}^{*}$ lie within the unit circle. Therefore, $\lim_{t \to
\infty} \| \tilde{\bar{\rho}}(t)^{[2]}\|= 0$, so that \[\lim_{t \to \infty}
\tilde{\bar{\rho}}(t)=\tilde{\bar{\rho}}_{\infty }^{T}=[\tilde{\bar{\rho}}(0)^{[1] T} 0 \cdots 0]^{T}.\]
Consequently,
\begin{equation}
 \bar{\rho}_{\infty }=T [\tilde{\bar{\rho}}(0)^{[1] T}  0 \cdots 0]^{T}=(i_{1} \pi_{1}+i_{2} \pi_{2}) \bar{\rho}(0).
\end{equation}
Having in mind the definition of $i_{1}$ and $i_{2}$, we conclude that $\bar{\rho}_{\infty 1 }=\bar{\rho}_{\infty 2 }= \cdots =
\bar{\rho}_{\infty n }$. \hspace*{\fill}\QED

Now we focus on the original recursion \eqref{compalg}. First we demonstrate the important fact, stemming from the structure of the matrices in (\ref{compalgm}), that the transformation $T$ from Lemma~4,  after being applied to the time-varying matrix $B(t)$, results in a matrix with the same structure as the transformed matrix $\bar{B}$ in \eqref{tminusone}.

\begin{lemma} \label{lemma:5} Matrix $B(t)$ in (\ref{compalg}) satisfies for all $t$
\begin{equation}
T^{-1} B(t) T=\left[ \begin{BMAT}{c.c}{c.c} 0_{2 \times 2} & 0_{2 \times (2n-2)}   \\ 0_{(2n-2) \times 2} &
B(t)^{*}
\end{BMAT} \right],
\end{equation}
where $B(t)^{*}$ is an $(2n-2) \times (2n-2)$ matrix and $T$ is given in Lemma~\ref{lemma:4}.
\end{lemma}
\par
The proof is given in the Appendix.

Now we are in a position to prove the following theorem dealing with the convergence of the main recursion \eqref{compalg} in the mean square sense and w.p.1.
\begin{theorem} \label{theorem:2} Let Assumptions A1), A2), A3) and A4) be satisfied. Then there exists $\delta''
> 0$ such that for all $\delta \leq \delta''$ in (\ref{compalg})
\begin{equation} \label{limitt}
 \lim_{t \to \infty} \hat{\rho}(t)= (i_{1} \pi_{1}+i_{2} \pi_{2}) \hat{\rho}(0)
\end{equation}
in the mean square sense and w.p.1. \end{theorem} {\bf Proof:} Using Lemmas~\ref{lemma:4} and \ref{lemma:5}, we define $\tilde{\rho}(t)=T^{-1} \hat{\rho}(t)$ and obtain
\begin{eqnarray} \label{onetwo1}
 \tilde{\rho}(t+1)^{[1]} & = &\tilde{\rho}(t)^{[1]};  \\
 \tilde{\rho}(t+1)^{[2]} &= &(I+\delta B(t)^{*}) \tilde{\rho}(t)^{[2]},  \nonumber
\end{eqnarray}
where ${\rm dim}\{ \tilde{\rho}(t)^{[1]}\}=2$, $
{\rm dim}\{\tilde{\rho}(t)^{[2]}\}=2n-2$. Recalling that $\bar{B}^{*}$
 is Hurwitz, we observe that there exists such a positive definite matrix $R^{*}$ that
\begin{equation} \label{q}
\bar{B}^{*T}R^{*}+R^{*}\bar{B}^{*}=-Q^{*},
\end{equation}
where $Q^{*}$ is positive definite. Define $q(t)= E \{\tilde{\rho}(t)^{[2]T} R^{*} \tilde{\rho}(t)^{[2]} \},$
and let $\lambda_{Q} =\min_{i} \lambda_{i}\{ Q^{*} \}$ and $k'=\max_{i}\lambda_{i}\{ E \{B(t)^{*} $ $
B(t)^{*T}\}\} $. From (\ref{onetwo1}) and A2) we obtain
\begin{equation}
q(t+1)= E\{\tilde{\rho}(t)^{[2]T} E\{ (I+\delta B(t)^{*})^{T} R^{*}  (I+\delta B(t)^{*}) \} \tilde{\rho}(t)^{[2]} \}
\end{equation}
and, further,
\begin{equation} \label{q1}
q(t+1) \leq \left[1 -  \delta \frac{ \lambda_{Q}}{\max_{i}\lambda_{i}\{R^{*}\}}+  \delta^2 k' \frac{\max_{i}
\lambda_{i}\{R^{*}\} }{ \min_{i} \lambda_{i} \{R^{*}\} } \right]q(t),
\end{equation}
having in mind that $E\{ B(t)^{*} \}=\bar{B}^{*}$. Consequently, there exists such a $\delta''$ that for $\delta
< \delta''$, $i=1, \ldots, n$, the term in the brackets at the right hand side of (\ref{q1}) is less than one.
Therefore, $q(t)$ tends to zero exponentially, implying that $\tilde{\rho}(t)^{[2]}$ converges to zero in the
mean square sense and with probability one. The result follows from $ \hat{\rho}(t)= T \tilde{\rho}(t)$, analogously with Theorem 1.
\hspace*{\fill}\QED

After clarifying the main structural properties of the algorithm, we shall extend the analysis to the more realistic case of correlated sequences $\{x(t)\}$. The results are especially important for the next section dealing with noisy measurements. We replace A2) with:

A2') Process $\{x(t) \}$ is weakly stationary with $E\{x(t)\} = \bar{x}$, $E\{x(t)x(t-d)\}=m(d)$, $m(0)=s^{2}$,
$ |x(t)| \leq K < \infty $ (w.p.1) and
\begin{align}
a)& \;\;\;\;\; |E \{ x(t)| \mathcal{F}_{t-\tau} \} - \bar{x}| =o(1), \;\; ({\rm w.p.1})
 \\ b)& \;\;\;\;\; |E \{ x(t)x(t-d)| \mathcal{F}_{t-\tau} \} - m(d)|=o(1), \;\; ({\rm w.p.1})
\end{align}
when $\tau \to \infty$, for any fixed $d \in \{0,1,2, \ldots\}$, $\tau > d$, where $\mathcal{F}_{t-\tau}$ denotes the minimal
$\sigma$-algebra generated by $\{ x(t-\tau), x(t-\tau-1), \ldots, x(0) \}$, and $o(1)$ denotes a function that
tends to zero as $\tau \to \infty$.

\begin{theorem} \label{theorem:3}  Let Assumptions A1), A2'), A3) and A4) be satisfied. Then there exists $\delta'''
> 0$ such that for all $\delta \leq \delta'''$ in (\ref{compalg})
$\lim_{t \to \infty} \hat{\rho}(t)= (i_{1} \pi_{1}+i_{2} \pi_{2}) \hat{\rho}(0)$ in the mean square sense and
w.p.1. \end{theorem} {\bf Proof:} Following the proof of Theorem~2, we first compute $\tilde{\rho}(i)=T^{-1} \hat{\rho}(t)$, and obtain
the same relations as in (\ref{onetwo1}). Iterating back the second relation $\tau$ time steps, one obtains
\begin{equation} \label{itback}
\tilde{\rho}(t+1)^{[2]}= \prod_{s=t}^{t-\tau} (I+ \delta B(s)^{*}) \tilde{\rho}(t-\tau)^{[2]}.
\end{equation}
Again, we define $q(t)= E \{\tilde{\rho}(t)^{[2]T} R^{*} \tilde{\rho}(t)^{[2]} \}$ and, after calculating $q(t+1)$ using (\ref{itback}), we extract the term linear in $\delta$
\begin{align}
& E \{\tilde{\rho}(t-\tau)^{[2]T}E \{ \sum_{s=t}^{t-\tau} (B(s)^{*T}R^{*}+R^{*} B(s)^{*})|\mathcal{F}_{t-\tau-1}
\} \tilde{\rho}(t-\tau)^{[2]} \}  \\& =E \{\tilde{\rho}(t-\tau)^{[2]T}[-(\tau+1)Q^{*} \nonumber \\ &\quad+E \{ \sum_{s=t}^{t-\tau}
(\tilde{B}(s)^{*T}R^{*}+R^{*} \tilde{B}(s)^{*})|\mathcal{F}_{t-\tau-1} \}] \tilde{\rho}(t-\tau)^{[2]} \},
\nonumber
\end{align}
where $t\geq \tau$, and we have written $B(t)^{*}$ as $B(t)^{*}=\bar{B}^{*}+\tilde{B}(t)^{*}$, with $E \{\tilde{B}(t)^{*}\}=0$.  According to A2') we can conclude the following:
\begin{align}
& |E \{ \tilde{\rho}(t-\tau)^{[2]T}E\{ \tilde{B}(s)^{*T}R^{*}+R^{*} \tilde{B}(s)^{*}|  \mathcal{F}_{t-\tau-1} \} \tilde{\rho}(t-\tau)^{[2]}
\}| \nonumber \\ & \leq  \phi(s-t+\tau+1) q(t-\tau),
\end{align}
for $t-\tau \leq s \leq t$, where $\phi(s) > 0$, $\lim_{s \to \infty} \phi(s)=0$. Since $ \lambda_{\min}(Q^{*}) > 0$ by definition, it is possible to find such $\tau_{0} > 0$ that for all $\tau \geq \tau_{0}$ and $t\geq \tau$
\begin{equation}
(\tau +1) \lambda_{\min}(Q^{*}) - \sum_{s=t}^{t-\tau} \phi(s-t+\tau+1) > \epsilon,
\end{equation}
where $\epsilon$ is a positive constant. Therefore, according to A2') there exist constants $k_{s}$, $s=2, \ldots, 2(\tau+1)$, such that
\begin{equation} \label{poly}
q(t+1) \leq (1- \epsilon\delta+ \sum_{s=2}^{2(\tau+1)} k_{s} \delta^{s}) q(t).
\end{equation}
 It follows from (\ref{poly}) that there exists a $\delta''' > 0$ such that $ 1- \epsilon \delta+ \sum_{s=2}^{2(\tau+1)} k_{s} \delta^{s} < 1$ for all $\delta < \delta'''$. The result
follows now in the same way as in the proof of Theorem~2. \hspace*{\fill}\QED
\par
\textbf{Remark 1} Stationarity of the random process $\{ x(t) \}$ assumed in A2) and A2') cannot be considered restrictive from the point of view of applications, since it encompasses a large variety of quickly and slowly varying real signals. This assumption is not essential for proving convergence of the proposed algorithm: it has been introduced primarily for the sake of focusing on the essential structural aspects of the algorithm and avoiding complex notation (especially in the next section where noise influence is analyzed). Notice, according to Lemmas 4 and 5, that a constant decoupling transformation $T$ can still be applied within the scope of the convergence analysis even when we have a time varying matrix $\bar{B}(t)$, owing to its specific structure; namely, we have $T^{-1} \bar{B}(t)T =\left[ \begin{BMAT}{c.c}{c.c} 0_{2 \times 2} & 0_{2 \times (2n-2)}   \\ 0_{(2n-2) \times 2} & \bar{B}(t)^{*} \end{BMAT} \right]$, where $\bar{B}(t)^{*}$ is Hurwitz and $T$ is obtained from $\bar{B}(t)$ for any selected $t=t'$. Moreover, notice that the conclusions of Theorem~1 hold, in general, provided the following unrestrictive condition holds: $\lim_{t} \prod_{\tau} (I- \bar{B}(t-\tau)^{*})=0$. 
\par 
Also, the applied methodology of convergence analysis, although adapted to the case of stationary signals, allows to conclude directly that the conclusions of the above theorems hold for changes of $\bar{B}(t)^{*}$ sufficiently slow. For example, in the proof of Theorem~2, if $R(t)^{*} > 0$ is a unique solution of $\bar{B}(t)^{*T} R(t)^{*}+R(t)^{*} \bar{B}(t)^{*}= -Q(t)$ for a preselected $Q(t) > 0$, then by defining $q(t)=  E \{\tilde{\rho}(t)^{[2]T} R(t)^{*} \tilde{\rho}(t)^{[2]} \}$ one obtains the results of Theorems 2 and 3 if $\|R(t+1)^{*}-R(t)^{*}\|$ is small enough. Moreover, it is not difficult to prove that the obtained results exactly hold when the signal is asymptotically stationary, \emph{i.e.}, when $\lim_{t \to \infty} E\{ x(t) \} = \bar{x}$ and $\lim_{t \to \infty} E\{x(t)x(t-d)\}=m(d)>0$. \hspace*{\fill}\QED

\section{Convergence Analysis: Communication Errors and Measurement Noise}

This section deals with the application of the proposed blind macro calibration methodology to \emph{lossy sensor networks}, including \emph{communication errors} and \emph{measurement noise}.
In the case of measurement noise, the main algorithm will undergo specific modifications based on introducing \emph{instrumental variables}, aimed at overcoming correlations preventing the original gradient scheme from converging to consensus.

\subsection{Communication Errors}
We assume that communication errors are manifested in two ways: 1) communication
dropouts and 2) additive communication noise. Accordingly, we formally assume:
\par
A5) The weights $\gamma_{ij}$ in the algorithm (\ref{alg}) are stochastic processes represented as $\{\gamma_{ij}(t)\}=\{u_{ij}(t) \gamma_{ij}\}$,
where $ \{u_{ij}(t)\}$ are i.i.d. binary random sequences,
such that $u_{ij}(t)=1$ with probability $p_{ij}$ ($p_{ij} > 0$ when $j \in \mathcal{N}_{i}$), and $u_{ij}(t)=0$ with
probability $1-p_{ij}$.
\par
A6) Instead of receiving $\hat{z}_{j}(t)$ from the $j$-th node, the $i$-th node receives
$\hat{z}_{j}(t)+\xi_{ij}(t)$, where $\{ \xi_{ij}(t) \}$ is an i.i.d. random sequence with $E \{ \xi_{ij}(t) \}=0$
and $E \{ \xi_{ij}(t)^{2} \}=(\sigma^{\xi }_{ij})^{2}<\infty$.
\par
A7) Processes $x(t)$, $u_{ij}(t)$ and $\xi_{ij}(t)$ are mutually independent.
\par
Denoting
\[ \nu_{i}(t)=   \sum_{j \in \mathcal{N}_{i}}\gamma_{ij}(t)\xi_{ij}(t)
 \left[ \begin{BMAT}{c}{cc} \alpha_{i} y_{i}(t) \\ 1 + \beta_{i}y_{i}(t) \end{BMAT} \right], \]
and $\nu(t)=\left[ \begin{BMAT}{ccc}{c} \nu_{1}(t) & \ldots & \nu_{n}(t) \end{BMAT} \right]$,  we obtain from
(\ref{compalg}) that
\begin{equation} \label{compalgn}
\hat{\rho}(t+1)=[I+ (\Delta(t) \otimes I_{2}) B'(t)] \hat{\rho}(t)+ \Delta(t) \nu(t),
\end{equation}
where $B'(t)= \Phi(t)( \Gamma(t) \otimes I_{2})$, with $\Gamma(t)$ obtained from $\Gamma$ by replacing constants
$\gamma_{ij}$ with time varying gains $\gamma_{ij}(t)$, according to A5).
\par
We will study convergence of the recursion (\ref{compalgn}) starting from the results of the previous section.
Notice first that, due to
mutual independence between the random variables in $B'(t)$, we have $E \{ B'(t) \}=\bar{B}'=  \bar{\Phi}(\bar{\Gamma} \otimes I_{2})$, where
$\bar{\Gamma}= E \{ \Gamma(t) \}$ is obtained from $\Gamma$ by replacing $\gamma_{ij}$ with $\gamma_{ij}p_{ij}$.
Defining $\tilde{B}'(t)= B'(t)-\bar{B}'$, we conclude that $ E\{ \tilde{B}'(t) \}=0$ and $E \{ \tilde{B}'(t)|
\mathcal{F}_{t-1} \}=0$. It is obvious that $\bar{B}'=
\bar{\Phi}(\bar{\Gamma} \otimes I_{2})$ has qualitatively the same properties as $\bar{B}$ in (\ref{compalgm}):
it has two eigenvalues at the origin and the remaining eigenvalues in the left half plane.
\par
Further, we now assume that the step sizes $\delta_i(t)$ satisfy the following assumption standard for stochastic approximation algorithms (\textit{e.g.}, \cite{hfchen}):

A8) $\delta_{i}(t)=\delta(t) > 0$, $\sum_{t=0}^{\infty} \delta(t) = \infty$, $\sum_{t=0}^{\infty} \delta(t)^{2} <
\infty$, $i=1, \ldots, n$.

Therefore, we have
\begin{equation} \label{noiserec}
\hat{\rho}(t+1)=(I+ \delta(t) \bar{B}') \hat{\rho}(t) + \delta(t) \tilde{B}'(t) \hat{\rho}(t)+ \delta(t) \nu(t).
\end{equation}

Let \[T'= \left[
\begin{BMAT}{c.c.c}{c} i_{1} & i_{2} & T'_{2n \times (2n-2)} \end{BMAT} \right], \] where
  $T'_{2n \times (2n-2)}$ is an $2n \times (2n-2)$
 matrix, such that ${\rm span}\{T'_{2n \times (2n-2) }\}={\rm span}\{\bar{B}'\}$. Then, $(T')^{-1}=\left[ \begin{BMAT}{c}{c.c.c} \pi'_{1} \\ \pi'_{2} \\ S'_{(2n-2) \times
2n} \end{BMAT} \right]$, where $\pi'_{1}$ and $\pi'_{2}$ are the left eigenvectors of $\bar{B}'$ corresponding
to the zero eigenvalue.

\begin{theorem} \label{theorem:4} Let Assumptions A2)--A8) be satisfied. Then, $ \hat{\rho}(t)$ generated by
(\ref{noiserec}) converges to $i_{1} w_{1}+i_{2} w_{2} $ in the mean square sense and w.p.1, where $w_{1}$ and
$w_{2}$ are scalar random variables satisfying $E\{w_1\}=\pi'_{1}\hat{\rho}(0)$ and $E\{w_2\}=\pi'_{2}\hat{\rho}(0)$. \end{theorem} {\bf Proof:}  Let $\tilde{\rho}(t)= [\tilde{\rho}_{1}(t) \;  \tilde{\rho}_{2}(t) \cdots
\tilde{\rho}_{2n}(t)]^{T}$ $=(T')^{-1} \hat{\rho}(t)$. Then, (\ref{noiserec}) gives
\begin{align} \label{onetwo2}
 \tilde{\rho}(t+1)^{[1]}=&\tilde{\rho}(t)^{[1]}+ \delta(t)G_{1}(t)\tilde{\rho}(t)+ \delta(t) \nu'(t),    \\
 \tilde{\rho}(t+1)^{[2]}=&(I+\delta(t) \bar{B}^{'*}) \tilde{\rho}(t)^{[2]} +\delta(t)G_{2}(t)\tilde{\rho}(t)+\delta(t) \nu''(t),
\end{align}
where $\tilde{\rho}(t)^{[1]}$ and $\tilde{\rho}(t)^{[2]}$ are defined as in (\ref{onetwo}), $\left[
\begin{BMAT}{c}{c.c} G_{1}(t) \\ G_{2}(t)  \end{BMAT} \right] = (T')^{-1} \tilde{B}'(t) T'$ in such a way that $G_{1}(t)$ contains
the first two rows, $ \nu'(t)=\left[ \begin{BMAT}{c}{c.c} \pi'_{1} \\ \pi'_{2}
\end{BMAT} \right] \nu(t) $ and $  \nu''(t)= S'_{(2n-2) \times 2n} \nu(t) $, while $\bar{B}^{'*}$ is a $(2n-2)
\times (2n-2)$ Hurwitz matrix such that $(T')^{-1} \bar{B}^{'} T'= {\rm diag} \{0_{2 \times 2} ,\bar{B}^{'*} \}$
(see Theorem 1). It is easy to verify that $E \{G_{1}(t)\}=0$ and $E \{G_{2}(t)\}=0$, as well as that $E\{
G_{1}(t)| \mathcal{F}_{t-1}\}=0$ and $E \{G_{2}(t)|\mathcal{F}_{t-1} \}=0$.
\par
Let $P^{*} > 0$ satisfy the Lyapunov equation $P^{*} \bar{B}^{'*} + \bar{B}^{'*T}P^{*}=-Q^{*}$ for some $Q^{*}
> 0$. Denote $s(t)=E\{ \| \tilde{\rho}(t)^{[1]} \|^{2} \}$ and $V(t)= E\{  \tilde{\rho}(t)^{[2]T}
P^{*} \tilde{\rho}(t)^{[2]} \}$. Then,
 directly following the methodology of \cite{huma3} (Theorem 11), one obtains
\begin{align} \label{sv}
 s(t+1)& \leq s(t)+C_{1}\delta(t)^{2}(1+s(t)+V(t))  \nonumber  \\
 V(t+1)& \leq (1-c_{0} \delta(t)) V(t) + C_{2} \delta(t)^{2}(1+s(t)+V(t)),
\end{align}
where $c_{0}$, $C_{1}$ and $C_{2}$ are appropriately chosen positive constants. According to \cite{huma3} (Lemma
12 and Theorem 11) and \cite{lizhang}, inequalities $(\ref{sv})$ give rise to the conclusion that $\tilde{\rho}(t)^{[1]}$ tends to
a vector random variable and $\tilde{\rho}(t)^{[2]}$ to zero in the mean square sense and
w.p.1. Finally, the claim of the theorem follows after calculating
 $ \lim_{t \to \infty} \hat{\rho}(t)= T'\left[ \begin{BMAT}{c}{c.c}
\lim_{t \to \infty} \tilde{\rho}(t)^{[1]} \\ 0  \end{BMAT} \right]$.
The expressions for $E\{w_1\}$ and $E\{w_2\}$ follow from Theorems 1 and 3. This is important property from the point of view of the global behaviour of the proposed calibration scheme.
\hspace*{\fill}\QED

\subsection{Measurement Noise: Instrumental Variables}
We assume in this subsection that the signal $x(t)$ is contaminated by additive measurement noise. This is formally defined by the following assumption:
\par
A9) Instead of $y_{i}(t)$  in (\ref{yi}), the sensors generate the signals contaminated by noise $y_{i}^{\eta}(t)=\alpha_{i} x(t) +
\beta_{i}+ \eta_{i}(t),$ where $ \{\eta_{i}(t) \}$, $i=1, \ldots n$, are zero mean i.i.d. random sequences with
$E\{ \eta_{i}(t)^{2} \} = (\sigma_{i}^{\eta})^{2}$, independent of the measured signal $x(t)$.
\par
Inserting
$y_{i}^{\eta}(t)$  instead of $y_{i}(t)$ in the basic algorithm (\ref{alg}), we obtain, after changing the
variables, the following ``noisy'' version of (\ref{algro}):
\begin{align} \label{algron}
 \hat{\rho}_{i}(t+1)=\hat{\rho}_{i}(t)+ \delta_{i}(t) \sum_{j \in \mathcal{N}_{i}}\gamma_{ij} \{& [\Phi_{i}(t)+
\Psi_{i}(t)]  [\hat{\rho}_{j}(t)-\hat{\rho}_{i}(t)] \nonumber \\&+
N_{ij}(t)\hat{\rho}_{j}(t)-N_{ii}(t)\hat{\rho}_{i}(t)\},
\end{align}
where $\Psi_{i}(t)= \eta_{i}(t)\left[
\begin{BMAT}{cc}{cc} \alpha_{i} x(t) & \alpha_{i} \\ \beta_{i} x(t) & \beta_{i}  \end{BMAT} \right],$
$
N_{ij}(t)=\frac{\eta_{j}(t)}{\alpha_{j}}\left[\begin{BMAT}{cc}{cc} \alpha_{i} y_{i}(t) & 0 \\ \beta_{i} y_{i}
(t) & 0
\end{BMAT} \right]+\left[\begin{BMAT}{cc}{cc} \frac{\eta_{j}(t)\eta_{i}(t)}{\alpha_{j}} & 0 \\ 0
 & 0
\end{BMAT} \right] $
and
$
N_{ii}(t)=\frac{\eta_{i}(t)}{\alpha_{i}}\left[\begin{BMAT}{cc}{cc} \alpha_{i} y_{i}(t) & 0 \\ \beta_{i} y_{i}
(t) & 0 \end{BMAT} \right]+\left[\begin{BMAT}{cc}{cc} \frac{\eta_{i}(t)^{2}}{\alpha_{i}} & 0 \\ 0
 & 0 \end{BMAT} \right]$, assuming $\alpha_i \neq 0$, $i=1,...,n$.
 Notice that $E \{\Psi_{i}(t) \}=0$, $E \{N_{ij}(t) \}=0$, but $E \{N_{ii}(t)\}=\left[\begin{BMAT}{cc}{cc}
\frac{(\sigma_{i}^{ \eta })^{2}}{\alpha_{i}} & 0 \\ 0
 & 0 \end{BMAT} \right] $.
 \par
 Assuming $\delta_{i}(t)=\delta(t)$, $i=1, \ldots, n$, we can write in accordance with (\ref{compalg})
\begin{equation} \label{compalgn1}
\hat{\rho}(t+1)=(I+ \delta(t)\{ [ \Phi(t)+ \Psi(t)](\Gamma \otimes I_{2}) + \tilde{N}(t)\} )\hat{\rho}(t),
\end{equation}
where  $ \Psi(t)= {\rm diag} \{\Psi_{1}(t), \ldots, \Psi_{n}(t) \}$ and $\tilde{N}(t) =[\tilde{N}_{ij}(t)]$ with
$ \tilde{N}_{ij}(t)=-\sum_{k, k \neq i}$ $ \gamma_{ik} N_{ii}(t)$ for $i=j$ and $ \tilde{N}_{ij}(t)= \gamma_{ij}
N_{ij}(t)$ for $i \neq j$, $i,j=1, \ldots, n$.
\par
Applying the methodology from the previous section to the analysis of (\ref{compalgn1}), we conclude that, instead of (\ref{compalgm}), we have
 now
$\bar{\rho}(t+1)=  [I+ \delta(t)(\bar{B}+\Sigma_{\eta})] \bar{\rho}(t)$,
where $\bar{B}$ is defined in (\ref{compalgm}) and $\Sigma_{\eta}=$ $ -{\rm diag} \{ \frac{(\sigma_{1}^{\eta
})^{2}}{\alpha_{1}} $ $ \sum_{j} \gamma_{1j}, 0, \ldots , $ $ \frac{(\sigma_{n}^{\eta })^{2}}{\alpha_{n}}
\sum_{j} \gamma_{nj}, 0 \}$. Under A2) and A3) the last recursion
 does not have the properties of (\ref{compalgm}), due to the additional term
$\Sigma_{\eta}$; the fact that all the row sums of $\bar{B}
+ \Sigma_{\eta}$ are now not equal to zero prevents the achievement of asymptotic consensus (see Theorem~1).

The above problem can be overcame in the case when the measurement noise variances $(\sigma_{i}^{\eta })^{2}$ are \textit{a priori} known. Consequently, the following algorithm, able to achieve asymptotic consensus, is proposed as a modification of \eqref{alg}:

\begin{align} \label{algn}
 \hat{\theta}_{i}(t+1)=\hat{\theta}_{i}(t)+ \delta(t)\{ \sum_{j \in \mathcal{N}_{i}}\gamma_{ij}
\epsilon^{\eta}_{ij}(t) \left[
\begin{BMAT}{c}{cc} y^{\eta}_{i}(t) \\ 1 \end{BMAT} \right] +\left[
\begin{BMAT}{c.c}{c.c} (\sigma_{i}^{\eta })^{2} \sum_{j \in \mathcal{N}_{i}}\gamma_{ij}& 0 \\ 0 & 0  \end{BMAT}
\right] \hat{\theta}_{i}(t)\},
\end{align}
where $\epsilon^{\eta}_{ij}(t)=\hat{z}^{\eta}_{j}(t)-\hat{z}^{\eta}_{i}(t)$ and $\hat{z}^{\eta}_{i}(t)=\hat{a}_{i}(t)
y^{\eta}_{i}(t)+\hat{b}_{i}(t)$, $i=1,\ldots,n$.

\begin{theorem} \label{theorem:5} Let Assumptions A2)--A4), A8) and A9) be satisfied.
Then, $\hat{\rho}(t)$  generated by (\ref{algn}) converges to $i_{1} w_{1}+i_{2} w_{2} $ in the mean square sense
and w.p.1, where $w_{1}$ and $w_{2}$ are scalar random variables satisfying $E\{w_1\}=\pi'_{1}\hat{\rho}(0)$ and $E\{w_2\}=\pi'_{2}\hat{\rho}(0)$. \end{theorem} {\bf Proof:} The algorithm (\ref{algn}) can be represented in the following compact form (compare with \eqref{compalgn1}):
\begin{equation} \label{compalgmnn}
\hat{\rho}(t+1)=\{ I+ \delta(t)[B(t)-\Sigma_{\eta}+\Psi(t)(\Gamma \otimes I_{2})+\tilde{N}(t)] \} \hat{\rho}(t).
\end{equation}
Since $B(t)=\bar{B}+\tilde{B}(t)$, where $\bar{B}$ is defined in (\ref{compalgm})
and $\tilde{B}(t)$ satisfies $E \{ \tilde{B}(t)\}=0 $ and is independent from $\tilde{B}(\tau)$, $\tau < t$, we
have
\begin{equation} \label{meas}
\hat{\rho}(t+1)=( I+ \delta(t)\bar{B}) \hat{\rho}(t)+\delta(t) E(t) \hat{\rho}(t),
\end{equation}
where $E(t)=\tilde{B}(t)-\Sigma_{\eta}+ \Psi(t)(\Gamma \otimes I_{2})+\tilde{N}(t)$ is a zero mean white noise
term. We observe that (\ref{meas}) represents structurally a special case of (\ref{noiserec}),
not containing the last, stochastic, term. Therefore, the proof can be completed by applying the  methodology of the proof of Theorem~4.  \hspace*{\fill}\QED

Notice that the above theorem was based on assumption A2): indeed, when both $\{ x(t) \} $ and $\{\eta_{i}(t) \}$  are white noise sequences, it is not surprising that the asymptotic consensus is achievable only
provided $ \sigma_{i}^{\eta } $, $i=1, \ldots, n$, are known. The more realistic assumption A2'), allowing correlated sequences $\{x(t)\}$ (treated in Section 2 in the noiseless case), allows overcoming the above mentioned  correlatedness problem without requiring any \emph{a priori} information.  The idea is to introduce \emph{instrumental variables} in the main algorithm in the way analogous to the one often used in the field of recursive methods for system identification, \emph{e.g.}, \cite{ljung,sods}. Instrumental variables have the basic property of being correlated with the measured signal, and uncorrelated with noise. If $\{\zeta_{i}(t)\}$ is the instrumental variable sequence of the $i$-th agent, one has to ensure that $\zeta_{i}(t)$ is correlated with $x(t)$ and uncorrelated with $\eta_{j}(t)$, $j=1, \ldots, n$. Under A2') a logical choice is to take the delayed sample of the measured signal as an instrumental variable, \emph{i.e.}, to take $\zeta_{i}(t)=y_{i}^{\eta}(t-d)$, where $d \geq 1$.  Consequently, we propose the following general \emph{calibration algorithm based on instrumental variables} able to cope with measurement noise:
\begin{equation} \label{alginst}
\hat{\theta}_{i}(t+1)=\hat{\theta}_{i}(t)+ \delta(t) \sum_{j \in \mathcal{N}_{i}}\gamma_{ij}
\epsilon_{ij}^{\eta}(t) \left[
\begin{BMAT}{c}{cc} y_{i}^{\eta}(t-d) \\ 1 \end{BMAT} \right],
\end{equation}
where  $d \geq 1$ and $\epsilon_{ij}^{\eta}(t)=\hat{z}_j^\eta(t)-\hat{z}_i^\eta(t)$, $\hat{z}_i^\eta(t)=\hat{a}_i(t) y_i^\eta(t) +\hat{b}_i(t)$, $i=1,\ldots,n$. Following the derivations from Section 2, one obtains from
(\ref{alginst}) the following relations involving explicitly $x(t)$ and the noise terms:
\begin{align} \label{algroninst}
\hat{\rho}_{i}(t+1)=\hat{\rho}_{i}(t)+ \delta(t) \sum_{j \in \mathcal{N}_{i}}\gamma_{ij} \{& (\Phi_{i}(t,d)+
\Psi_{i}(t,d))(\hat{\rho}_{j}(t)-\hat{\rho}_{i}(t))  \nonumber \\
 &+ N_{ij}(t,d)\hat{\rho}_{j}(t)-N_{ii}(t,d)\hat{\rho}_{i}(t)\},
\end{align}
where \[ \Phi_{i}(t,d)=   \left[
\begin{BMAT}{cc}{cc} \alpha_{i}\beta_{i}x(t)+ \alpha_{i}^{2}x(t)x(t-d)  &  \alpha_{i}\beta_{i}+  \alpha_{i}^2 x(t-d) \\
 (1+\beta_{i}^2)x(t) + \alpha_{i}\beta_{i}x(t)x(t-d) & 1 + \beta_{i}^2+\alpha_{i}\beta_{i}x(t-d) \end{BMAT}
\right],\]
 \[ \Psi_{i}(t,d)= \eta_{i}(t-d)\left[
\begin{BMAT}{cc}{cc} \alpha_{i} x(t) & \alpha_{i} \\ \beta_{i} x(t) & \beta_{i}  \end{BMAT} \right],\]
\[
N_{ij}(t,d)=\frac{\eta_{j}(t)}{\alpha_{j}}\left[\begin{BMAT}{cc}{cc} \alpha_{i} y_{i}(t-d) & 0 \\ \beta_{i}
y_{i} (t-d) & 0
\end{BMAT} \right]+\left[\begin{BMAT}{cc}{cc} \frac{\eta_{j}(t)\eta_{i}(t-d)}{\alpha_{j}} & 0 \\ 0
 & 0
\end{BMAT} \right] \]
and
\[
N_{ii}(t,d)=\frac{\eta_{i}(t)}{\alpha_{i}}\left[\begin{BMAT}{cc}{cc} \alpha_{i} y_{i}(t-d) & 0 \\ \beta_{i}
y_{i} (t-d) & 0 \end{BMAT} \right]+\left[\begin{BMAT}{cc}{cc} \frac{\eta_{i}(t)\eta_{i}(t-d)}{\alpha_{i}} & 0 \\
0
 & 0 \end{BMAT} \right]. \]
In the same way as in (\ref{compalgn1}), we have
\begin{equation} \label{compalgnd}
\hat{\rho}(t+1)=(I+ \delta(t)\{ [ \Phi(t,d)+ \Psi(t,d)](\Gamma \otimes I_{2}) + \tilde{N}(t,d)\} )
\hat{\rho}(t),
\end{equation}
where $ \Phi(t,d)= {\rm diag} \{\Phi_{1}(t,d), \ldots, \Phi_{n}(t,d) \}$, $ \Psi(t,d)= {\rm diag}
\{\Psi_{1}(t,d), $ $ \ldots, $ $\Psi_{n}(t,d) \}$, $\tilde{N}(t,d) =[\tilde{N}_{ij}(t,d)]$, where $
\tilde{N}_{ij}(t,d)=-\sum_{k, k \neq i}$ $ \gamma_{ik} N_{ii}(t,d)$ for $i=j$ and $ \tilde{N}_{ij}(t,d)=
\gamma_{ij} N_{ij}(t,d)$ for $i \neq j$, $i,j=1, \ldots, n$.
\par
In order to formulate a convergence theorem for \eqref{compalgn1}, we need the following modification of
A4):
\par
A4') $m(d) > \bar{x}^{2}$ for some $d=d_{0} \geq 1$,
\par
implying that the correlation $m(d_0)$ should be large enough. Similarly as in the case of A4), we conclude that A4') implies that $-\bar{\Phi}(d)=-E \{ \Phi_{i}(t,d) \}$ is Hurwitz. 

\par
\begin{theorem} \label{theorem:6} Let Assumptions A2'), A3), A4'), A8) and A9) be satisfied. Then $\hat{\rho}(t)$ generated by
(\ref{compalgnd}) with $d=d_{0}$ converges to $ i_{1}w_{1}+i_{2} w_{2}$ in the mean square sense and w.p.1,
where $w_{1}$ and $w_{2}$ are scalar random variables satisfying $E\{w_1\}=\pi'_{1}\hat{\rho}(0)$ and $E\{w_2\}=\pi'_{2}\hat{\rho}(0)$. \end{theorem} {\bf Proof:} The proof starts from the demonstration that
\begin{equation} \label{tran}
 T^{-1} B(t,d) T= {\rm diag} \{ 0_{2 \times 2},
B(t,d)^{*} \},
\end{equation}
where $T$ is defined in Lemma~5, $B(t,d)=\Phi(t,d)(\Gamma \otimes I_{2})$, and $B(t,d)^{*}$
is an $(2n-2) \times (2n-2)$ matrix. In analogy with (\ref{w1b}) and (\ref{w2b}), we can write expressions for the components
of $v(t,d)=w^{T}B(t,d)$, where $w$ is a left eigenvector of $\bar{B}$ corresponding to the zero eigenvalue.
Using Lemma 6, it is immediately clear that  $v_{2i-1}(t,d)=0$ and  $v_{2i}(t,d)=0$, $i=1,
\ldots, n$, so that (\ref{tran}) directly follows.

We now compute  $\tilde{\rho}(i)=T^{-1} \hat{\rho}(t)$ and
obtain that
\begin{align}
  \tilde{\rho}(t+1)^{[1]}  = &\tilde{\rho}(t)^{[1]}+\delta(t) H_{1}(t,d) \tilde{\rho}(t) \label{onetwo11}  \\
  \tilde{\rho}(t+1)^{[2]} = &(I+\delta(t) B(t,d)^{*}) \tilde{\rho}(t)^{[2]} +\delta(t) H_{2}(t,d) \tilde{\rho}(t), \label{onetwo12}
\end{align}
where $\tilde{\rho}(t)^{[1]}=[\tilde{\rho}_{1}(t)
 \;\; \tilde{\rho}_{2}(t)]^{T}$, $
\tilde{\rho}(t)^{[2]}=[\tilde{\rho}_{3}(t) \cdots \tilde{\rho}_{2n}(t)]^{T}$, $H(t,d)=\left[
\begin{BMAT}{c}{c.c} H_{1}(t,d) \\ H_{2}(t,d)  \end{BMAT} \right]$\hspace{10mm} $ = T^{-1}[\Psi_{1}(t,d)(\Gamma \otimes I_{2})+
\tilde{N}(t,d)]T$, so that $H_{1}(t,d)$ contains the first two rows; notice that $\{ H(t,d) \}$ is i.i.d. and zero
mean, with $E\{ H(t,d)|\mathcal{F}_{t-1}\}=0 $.
\par
After iterating back the relation in (\ref{onetwo12}) $\tau$ times, one obtains
\begin{align} \label{ite}
  \tilde{\rho}(t+1)^{[2]}=&\Pi(t,t-\tau,d)\tilde{\rho}(t-\tau)^{[2]} +\sum_{\sigma=t-\tau}^{t}
 \Pi(t,\sigma+1,d)\delta(\sigma)H_{2}(\sigma,d) \tilde{\rho}(\sigma),
\end{align}
where $\Pi(t,s,d)=\prod_{\sigma=s}^{t} (I+ \delta(\sigma) B(\sigma,d)^{*})$, with $\Pi(t,t+1,d)=I$. \par Having
in mind A4'), we conclude that $\bar{B}(d)^{*}=E \{B(t,d) \}$ is Hurwitz for $d=d_{0}$; therefore, there exist
such symmetric positive definite matrices $R^{*}$ and $Q^{*}$ that $R^{*} \bar{B}(d_{0})^{*} +
\bar{B}(d_{0})^{*T}R^{*}=-Q^{*}$. Denote $s(t)=E\{ \| \tilde{\rho}(t)^{[1]} \|^{2} \}$ and $V(t)= E\{
\tilde{\rho}(t)^{[2]T} R^{*} \tilde{\rho}(t)^{[2]} \}$, as in the proof of Theorem~4. Calculation of $V(t)$ from (\ref{ite})
is straightforward, because $E \{ H_{2}(t,d) \tilde{\rho}(\sigma)\}=0$ for all $\sigma$. The crucial
term in the final expression is the linear part with respect to $\delta(\sigma)$, $\sigma=\{t-\tau, \ldots, t\}$, of $E \{\tilde{\rho}(t-\tau)^{[2]T} \Pi(t,t-\tau,d)^{T}
\Pi(t,t-\tau,d) \tilde{\rho}(t-\tau)^{[2]} \}$,
analogously with the case of time-invariant gains $\delta$ in Theorem~5. According to A2'), we obtain
\begin{align}
 &E \{ \tilde{\rho}(t-\tau)^{[2]T}E \{ R^{*} \tilde{B}(\sigma,d_{0})^{*}+ \tilde{B}(\sigma,d_{0})^{*T}R^{*}|\mathcal{F}_{t-\tau-1} \} \tilde{\rho}(t-\tau)^{[2]} \} \nonumber \\ & \leq \varphi(\sigma-t+\tau) V(t-\tau),
\end{align}
$t-\tau \leq \sigma \leq t$, where $\varphi(t) > 0$ and  $\lim_{t \to \infty} \varphi(t)=0$. Therefore, there exists $\tau_{0} > 0$ such that for all $ \tau \geq \tau_{0}$, $t \geq \tau$
\begin{equation}
\lambda_{\min}(Q^{*}) \sum_{\sigma=t-\tau}^{t} \delta(\sigma)  - \sum_{\sigma=t-\tau}^{t}
\varphi(\sigma-t+\tau)\delta(\sigma)
> \epsilon,
\end{equation}
for some $\epsilon>0$, since $ \lambda_{\min}(Q^{*}) > 0$ by definition. Therefore, for $t$ large enough, we have
\begin{align} \label{V}
 V(t+1) \leq (1- c_{0}\delta(t)) V(t-\tau) +C_{1} \sum_{\sigma=t-\tau}^{t} \delta(\sigma)^{2} (1+s(\sigma)+V(\sigma)),
\end{align}
where $c_{0} > 0$ and $C_{1} > 0$ are generic constants. Since from the first relation in (\ref{onetwo11}) we
have directly
\begin{equation} \label{s}
s(t+1) \leq s(t)+C_{1}\delta(t)^{2}(1+s(t)+V(t)),
\end{equation}
having in mind that $E\{ H_{1}(t,d)|\mathcal{F}_{t-1} \}=0$, recursions (\ref{V}) and (\ref{s}) can be treated
like the recursions in (\ref{sv}), giving
rise to the
conclusion that $\tilde{\rho}(t)^{[1]}$ tends to a vector random variable and $\tilde{\rho}(t)^{[2]}$ tends to zero in the mean square sense and
w.p.1. The result follows by calculating $ \lim_{t \to \infty} \hat{\rho}(t)= T\left[ \begin{BMAT}{c}{c.c}
\lim_{t \to \infty} \tilde{\rho}(t)^{[1]} \\ 0  \end{BMAT} \right]$.  \hspace*{\fill}\QED
\par
\textbf{Remark 2} After presenting the most important aspects of the proposed blind calibration methodology encompassing both noiseless and noisy environments, we are in the position to make a few comments on the problem of its relationship with the algorithms for \emph{time synchronization} in sensor networks, which has attracted a lot of attention (\emph{e.g.}, \cite{giku,sowa,scfi,jichen,ccsz,liba} and the references therein). Indeed, after coming back to the main measurement model, one easily realizes that in the case of time synchronization one has the form (\ref{yi}) with the absolute time $t$ replacing the signal value $x(t)$. An insight into the developed methodologies for time synchronization of sensor networks, motivated by the idea to evaluate possibilities of their potential application to the posed problem of blind calibration and to make comparisons with the proposed algorithm, leads to the conclusion that the time synchronization scheme proposed in \cite{scfi} could be considered to be formally and essentially the closest to the above proposed algorithm. The estimation scheme proposed in \cite{scfi} is, indeed, related to the estimation of the parameters of the calibration functions (\ref{calfun}) based on the utilization of local time measurements, but, however, consists of one separate recursion for the relative drift estimation and one  separate recursion for the estimation of offsets, relying on the obtained relative drifts. Trying to reformulate this scheme in the light of the calibration problem and the proposed methodology, one can start from the time difference model $\Delta y_{i}(t)=y_{i}(t+1)-y_{i}(t)=\alpha_{i} \Delta x(t)$, where $\Delta x(t)=x(t+1)-x(t)$, and construct a gradient recursion for $\hat{a}_{i}$ following the proposed methodology, having in mind that $\Delta y_{i}(t)$ does not depend on $\beta_{i}$. On the other hand, the estimation of $b_{i}$  has to start from (\ref{yi}); it has the form of the recursion for $\hat{b}_{i}$ in (\ref{alg}), but should use $\hat{a}_{i}$ generated by first recursion. Such a combined gradient algorithm based on utilization of $\Delta y_{i}(t)$ resembles the algorithm from \cite{scfi}; it is possible to expect that it has similar behavior as the proposed scheme in the noiseless case. However, the authors in \cite{scfi} do not consider the case of stochastic disturbances. This case could eventually be conceived using the ideas from Section 4, but it requires both theoretical and practical justification.
\par
In general, it is important to observe that the introduction of $t$ instead of $x(t)$ in the basic relation (\ref{yi}) suffers from the very basic problem that unboundedness of the linear function contradicts the requirements for boundedness of the second order moments of $x(t)$ (typical for stochastic approximation algorithms) and that it is not possible to guarantee convergence of the obtained recursions. This indicates that formal transfers of methodologies from one domain to the other should be done with extreme caution. \hspace*{\fill}\QED

\subsection{Convergence Rate}

A closer view on the asymptotic formulae in the above theorems shows that the asymptotic convergence rate of all the analyzed algorithms follows general statements related to stochastic approximation algorithms. Focusing the attention on the basic aspects of \emph{convergence to consensus} determined by the behavior of $\tilde{\rho}(t)^{[2]}$ in the context of all the above theorems, we have the following result giving an upper bound of the mean square error:
\par
\begin{corollary} Under the assumptions of any of the Theorems 4-6, together with
\par
A2") $\lim_{t \to \infty} (\delta(t+1)^{-1}-\delta(t)^{-1})= d \geq 0$,

\noindent
there exists such a positive $\sigma' < 1$ that for all $0 < \sigma < \sigma'$ the asymptotic consensus is achieved by the analyzed algorithms with the convergence rate
\begin{equation}
E\{\|\tilde{\rho}(t)^{[2]}\|^{2} \}= o(\delta(t)^{\sigma}).
\end{equation}
when $t \rightarrow \infty$. \end{corollary}
{\bf Proof:} Select the pair of inequalities related to $s(t)$ and $V(t)$ in (\ref{sv}) (the same methodology can be applied to (\ref{s}) and (\ref{V})). Applying Lemma~12 from \cite{huma3}, one obtains
 \begin{equation} \label{rate}
V(t+1) \leq (1-c_{0} \delta(t))V(t) + C_{1} \delta(t)^{2},
\end{equation}
where $c_{0}$ and $C_{1}$ are generic constants.
\par
From A2') and A2") we have by Taylor expansion
\begin{equation}
 \frac{\delta(t)^{\sigma}}{\delta(t+1)^{\sigma}} = 1+ \sigma \frac{\delta(t)-\delta(t+1)}{\delta(t+1)} + O((\frac{\delta(t)-\delta(t+1)}{\delta(t+1})^{2}),
\end{equation}
 having in mind that $ \frac{\delta(t)-\delta(t+1)}{\delta(t+1)} \to 0$.
Consequently, from (\ref{rate}) we have
\begin{equation}
 \frac{V(t+1)}{\delta(t+1)^{\sigma}} \leq \frac{\delta(t)^{\sigma}}{\delta(t+1)^{\sigma}}
 [1+c_{0} \delta(t)] \frac{V(t)}{\delta(t)^{\sigma}} + C_{1} \delta(t)^{2-\sigma},
 \end{equation}
 so that
\begin{align}
 W(t+1) \leq &\{1+[-c_{0}+\sigma \frac{\delta(t)-\delta(t+1)}{\delta(t+1)\delta(t)}+ \sigma \frac{\delta(t)-\delta(t+1)}{\delta(t+1)}  \nonumber \\
&+O((\frac{\delta(t)-\delta(t+1)}{\delta(t+1})^{2})] \delta(t) \} W(t) + C_{1} \delta(t)^{2-\sigma}
\end{align}
where $W(t)=  \frac{V(t)}{\delta(t)^{\sigma}}$. Taking into account A2"), for $t$ large enough one has
\begin{equation}
W(t+1) \leq [1+(-c_{0}+\sigma d +\varepsilon) \delta(t)]W(t) + C_{1} \delta(t)^{2-\sigma}.
\end{equation}
Having in mind that $c_{0} > 0$, it is evident that there exists $\sigma' > 0$ such that $c_{1}=-c_{0}+\sigma d +\varepsilon$ is negative for all $0 < \sigma \leq \sigma'$, since $\varepsilon$ can be made arbitrarily small. Therefore, according to the well known results (\emph{e.g.}, \cite{pol2,hfchen}),  $\lim_{t \to \infty} W(t)=0$, and the assertion immediately follows. \hspace*{\fill}\QED
\par
In practice it is difficult to make an \emph{a priori} estimate of the value of $\sigma'$. It is, however, clear that it depends directly on $ \lambda_{\min}(Q^{*})$, which, in turn, depends on the sensor and network properties expressed by matrix $B(t)$ (or $B(t,d)$) through the corresponding Lyapunov equations (see \emph{e.g.}, (\ref{q})). It is possible to choose $Q^{*} > 0$ and obtain uniquely $P^{*} > 0$ as a consequence of the fact that the nonzero eigenvalues of $\bar{B}$ (or $\bar{B(d)}$) are in the left half plane. Without going into details of the relationship between the eigenvalues of $Q^{*}$,  $P^{*}$ and $\bar{B}^{*}$ in general, we can notice here that in the case of undirected graphs and symmetric matrices $\bar{B}$, $c_{0}$ is proportional to the largest eigenvalue of $\bar{B}$. This fact leads to the general qualitative conclusion that the convergence rate of the proposed algorithms depends on the properties of the underlying graph, including its connectivity (see \cite{gr}). The number of nodes increases dimensionality of the parameter estimates and potentially decreases convergence rate in the same way as in stochastic approximation schemes; however, convergence rate increases by increasing connectedness of the graph. In the case when the graph is fully connected, convergence rate is high at the expense of a large number of direct communication links; a compromise should be found, as in all analogous problems in wireless sensor networks.
\par
In general, according to \cite{bm}, the choice of $\delta(t)$ should be based on the following qualitative estimate
\begin{equation} \label{ratebm}
E\{\|\tilde{\rho}(t)^{[2]}\|^{2} \} \leq v_{1}(\delta(t))+v_{2}(\delta(t))\exp\{-k_{1}\sum_{\tau} \delta(t-\tau)\},
\end{equation}
where $v_{1}(\delta(t)) \to 0$ following $\delta(t)$, $v_{2}(\cdot)$ is bounded and $k_{1} > 0$. The first term in (\ref{ratebm}) depends on the noise and the second on the initial conditions. Obviously, the choice of $\delta(t)$ should be based on a compromise between these two terms. Assuming that $\delta(t)$ has the standard form $\delta(t)=m_{1}/(m_{2}+t^{\mu})$, $m_{1},m_{2} > 0$, $\frac{1}{2} < \mu \leq 1$, the values of $\mu$ closer to $\frac{1}{2}$ would give faster convergence at the expense of noise immunity; the values of $\mu$ closer to one provide the opposite effect. In the noiseless case, the choice $\delta(t)=\delta={\rm const}$ provides an exponential convergence rate, since $v_{1}(\cdot)=0$.
\par
Furthermore, it is possible to formulate the proposed algorithm (\ref{alg}) such that $\delta_{i}(t)$ are $2 \times 2$ matrices defined using, \textit{e.g.}, the least square methodology, expecting higher convergence rate. It remains as a future work to clarify the convergence conditions of such an algorithm.
\par

\section{Macro Calibration Including Sensors with Fixed Characteristics}

According to Section 2, the choice of the elements of the matrix $\Gamma$, dictated by the relative importance (precision) of sensors in a given network, plays an important role in achieving good performance of the proposed method in practice. As mentioned earlier it is possible to increase importance of the sensors from a given set $\mathcal{N}^{f}$ by multiplying all $\gamma_{ij}$, $i \in \mathcal{N}^{f}$, $j=1,...,n$, $j \neq i$, with a small positive number. In such a way the corresponding $\hat{\rho}_i(t)$ remains closer to its initial condition $\hat{\rho}_i(0)$, $i \in \mathcal{N}^{f}$, thus having more influence on the global point of convergence. In the limit, sensors from $\mathcal{N}^{f}$ can be left unchanged (with fixed characteristics), so that the recursions (\ref{alg}) are applied only to the nodes $i \in \mathcal{N}-\mathcal{N}^{f}$. \par Take as an  example the case when one of the sensors, say $k$-th, $k \in \{1, \ldots, n
\}$, is taken as a reference because it has ideal (or desirable) characteristics; then, the whole calibration
network can be ``pinned'' to that sensor. The proposed algorithm (\ref{alg}) can be simply applied by setting
\begin{equation} \label{pin}
\hat{\theta}_{k}(t+1)=\hat{\theta}_{k}(t)
\end{equation}
where the initial condition $\hat{\theta}_{k}(0)=\hat{\theta}_{k0}$ should be appropriately chosen since the whole
calibration algorithm should ensure convergence of $\hat{\rho}_{i}(t)$, $i=1, \ldots, n, i \neq k$, to
the same vector $\hat{\rho}_{k0}=\left[
\begin{BMAT}{cc}{cc} \alpha_{k}  &  0  \\\beta_{k}& 1 \end{BMAT} \right] \hat{\theta}_{k0}$ (in the ideal case $\hat{\rho}_{k0} =\left[
\begin{BMAT}{c}{cc} 1 \\  0 \end{BMAT} \right])$.
The corresponding modification of the general form of
the algorithm consists of setting to zero all the block matrices in the $k$-th block row of $B(t)$ in
(\ref{compalg}), \textit{e.g.}, by setting $\gamma_{kj}=0$, $j=1,\ldots,n$.
An insight into assumptions A1) -- A8) shows that the only new aspect resulting from the application of (\ref{pin}) is a change in the structure of the graph $\mathcal{G}$, since the arcs leading to the node $k$ are eliminated by (\ref{pin}). However, A3) still holds, since the new graph still has a spanning tree with the node $k$ as the center node. Therefore, $\hat{\rho}_{i}(t)$, $i=1, \ldots, n$, converge in the sense of all the above proven theorems to the same limit. Since $ \lim_{t \to \infty} \hat{\rho}_{k}(t) = \hat{\rho}_{k0}$ by definition, this limit is equal to $\hat{\rho}_{k0}$.
\par
In general, the situation is more complex. Assume that $\mathcal{N}^{f}=\{k_{1},$ $ \ldots, k_{|\mathcal{N}^{f}|} \}$ is the subset of ``leaders'', \emph{i.e.}, of
sensors with arbitrary fixed characteristics $\rho^{f}_{k}=\left[
\begin{BMAT}{c}{cc} g^{f}_{k} \\  f^{f}_{k} \end{BMAT}
\right]$, $k \in \mathcal{N}^{f}$; let $\tilde{\rho}^{f}=[\rho_{k_{1}}^{fT} \cdots $ $ \rho_{k_{|\mathcal{N}^{f}|}}^{fT}]^{T}$. The above proposed algorithms can be applied in this situation in
such a way as to introduce (\ref{pin}) for all $k \in \mathcal{N}^{f}$. Let $\mathcal{N}-\mathcal{N}^{f}=$ $\{ l_{1}, \ldots, l_{|\mathcal{N}-\mathcal{N}^{f}|} \}$ and let $\hat{\rho}^{f}(t)=[\hat{\rho}_{l_{1}}(t)^{T} \cdots $ $ \hat{\rho}_{l_{|\mathcal{N}-\mathcal{N}^{f}|}}(t)^{T}]^{T}$ represent the vector of all the parameters to be adjusted. In the situation when $|\mathcal{N}^{f}| > 1$, the above conclusions related to the graph structure for $|\mathcal{N}^{f}| = 1$ do not hold any more; namely, it is easy to see that the resulting graph does not satisfy A3).
The next theorem treats convergence of the basic algorithm (\ref{alg}) in the case of arbitrary $\mathcal{N}^{f}$.
\par
\begin{theorem} \label{theorem:7} Let Assumptions A1), A2) and A4) be
satisfied and let all the nodes from $\mathcal{N} - \mathcal{N}^{f}$ be reachable from all the nodes in $\mathcal{N}^{f}$.
Then there exists $\delta''
> 0$ such that for all $\delta \leq \delta''$, the algorithm (\ref{alg}) combined with (\ref{pin})
for all $k \in \mathcal{N}^{f}$ provides convergence in the mean square sense and w.p.1 of $\hat{\rho}^{f}(t)$ to the limit
\begin{equation} \label{limit}
\hat{\rho}^{f}=-(\Gamma^{f} \otimes I_{2})^{-1} (\tilde{\Gamma}^{f} \otimes I_{2}) \tilde{\rho}^{f},
 \end{equation}
 where $\Gamma^{f}$ is matrix obtained from $\Gamma$ by deleting its rows and columns with indices $k_{1}, \ldots, k_{|\mathcal{N}^{f}|}$, \emph{i.e.}
\[\Gamma^{f} =\left[
\begin{BMAT}{cccc}{cccc} -\sum_{j=1, j \neq l_{1}}^{n} \gamma_{l_{1}j}  &  \gamma_{l_{1} l_{2}} & \cdots & \gamma_{l_{1} l_{|\mathcal{N}-\mathcal{N}^{f}|}} \\ \gamma_{l_{2} l_{1}} & -\sum_{j=1, j \neq l_{2}}^{n}
\gamma_{l_{2}j} & \cdots & \gamma_{l_{2}l_{|\mathcal{N}-\mathcal{N}^{f}|}} \\ & & \ddots & \\ \gamma_{l_{|\mathcal{N}-\mathcal{N}^{f}|} l_{1}} & \gamma_{l_{|\mathcal{N}-\mathcal{N}^{f}|} l_{2}} & \cdots & -\sum_{j=1, j \neq l_{|\mathcal{N}-\mathcal{N}^{f}|}}^{n}
\gamma_{l_{|\mathcal{N}-\mathcal{N}^{f}|}j}  \end{BMAT} \right].\]
while \[\tilde{\Gamma}^{f}=\left[ \begin{BMAT}{ccc}{ccc} \gamma_{l_{1}k_{1}} & \cdots & \gamma_{l_{1}k_{|\mathcal{N}^{f}|}} \\ & \cdots & \\ \gamma_{l_{|\mathcal{N}-\mathcal{N}^{f}|} k_{1}} & \cdots & \gamma_{l_{|\mathcal{N}-\mathcal{N}^{f}|} k_{|\mathcal{N}^{f}|} } \end{BMAT} \right]\]. \end{theorem} {\bf Proof:} Coming back to (\ref{algro}), one obtains
\begin{align}
\hat{\rho}_{i}(t+1)= &\hat{\rho}_{i}(t)+
\delta \sum_{j \in (\mathcal{N}_{i}-\mathcal{N}^{f})} \gamma_{ij} \Phi_{i}(t)
 [\hat{\rho}_{j}(t)-\hat{\rho}_{i}(t)]
\nonumber \\ &+\delta \sum_{j \in \mathcal{N}_{i} \cap \mathcal{N}^{f}} \gamma_{ij}\Phi_{i}(t)[\rho_{j}^{f}  -\hat{\rho}_{i}(t)],
\end{align}
so that we obtain
\begin{equation} \label{stpoint}
\hat{\rho}^{f}(t+1)=[I+ \delta \Phi^{f}(t) (\Gamma^{f} \otimes I_{2}) ]\hat{\rho}^{f}(t) +\delta \Phi^{f}(t) (\tilde{\Gamma}^{f} \otimes I_{2}) \tilde{\rho}^{f}
\end{equation}
where $\Phi^f(t)$ is obtained from $\Phi(t)$ by deleting its rows and columns with indices $k_{1}, \ldots, k_{|\mathcal{N}^{f}|}$.
It is clear that the stationary point of the recursion (\ref{stpoint}) is $\hat{\rho}^{f}=-(\Gamma^{f} \otimes I_{2})^{-1} (\tilde{\Gamma}^{f} \otimes I_{2}) \tilde{\rho}^{f}$, having in mind that $\Gamma^{f}$ is an M-matrix (according to Lemma~4). Therefore, we have
\begin{equation} \label{compf}
r^{f}(t+1)=(I+  \delta\Phi^{f}(t)( \Gamma^{f} \otimes I_{2}) ) r^{f}(t)
\end{equation}
where $r^{f}(t)=\hat{\rho}^{f}(t)-\hat{\rho}^{f}$. According to Lemma 4, $\Phi^{f}(t)( \Gamma^{f} \otimes I_{2})$ is Hurwitz.
The last relation allows direct application of Theorem 2. Thus the result.
\hspace*{\fill}\QED

It is possible to demonstrate directly from (\ref{stpoint}) that in the case when all the parameters in $\mathcal{N}^{f}$ are equal $\rho^{f}_{k}=\rho^{f}$, the rest of the sensor parameters will converge to $\rho^{f}$.

It is now straightforward to combine the result from the previous subsection with the results of Theorem~8, and prove convergence of the algorithm \eqref{compalgnd} in the presence of uncertainty in the form of both communication errors and measurement noise.

\begin{theorem} \label{theorem:8} Let Assumptions A2'), A4') and A5)--A9) be satisfied and let all the nodes from $\mathcal{N} - \mathcal{N}^{f}$ be reachable from all the nodes in $\mathcal{N}^{f}$.
 Then, the algorithm (\ref{alginst}) in which $\gamma_{ij}$ is replaced by $\gamma_{ij}(t)$ which satisfies A7), $\epsilon^{\eta}_{ij}(t)=\hat{z}^{\eta}_{j}(t)-\hat{z}^{\eta}_{i}(t) + \xi_{ij}(t)$ and
$\hat{z}^{\eta}_{i}(t)=\hat{a}_{i}(t) y^{\eta}_{i}(t)+\hat{b}_{i}(t)$, provides, combined with (\ref{pin}) for all nodes in $\mathcal{N}^{f}$, convergence in the mean square sense and w.p.1. of $\hat{\rho}^{f}(t)$ to the limit defined by (\ref{limit}). \end{theorem} {\bf Proof:} Starting from (\ref{compalgnd}), one obtains using the notation introduced in the previous theorem
\begin{align} \label{rin}
\hat{\rho}_{i}(t+1)=&\hat{\rho}_{i}(t)+\delta(t) \sum_{j \in (\mathcal{N}_{i}-\mathcal{N}^{f})}\gamma_{ij}(t) \{[ \Phi_{i}(t,d) +
\Psi_{i}(t,d)](\hat{\rho}_{j}(t) - \hat{\rho}_{i}(t))   \nonumber \\ & +N_{ij}(t,d) \hat{\rho}_{j}(t)-N_{ii}(t,d) \hat{\rho}_{i}(t) \}
  \nonumber \\ &+  \delta(t) \sum_{j \in \mathcal{N}_{i} \cap \mathcal{N}^{f}}\gamma_{ij}(t)\{ [ \Phi_{i}(t,d) +
\Psi_{i}(t,d)] (\rho_{j}^{f}- \hat{\rho}_{i}(t))  \nonumber \\&-N_{ii}(t,d) \hat{\rho}_{i}(t)  \} +\delta(t)\nu_{i}(t,d)
\end{align}
where $\nu_{i}(t,d)=   \sum_{j \in \mathcal{N}_{i}}\gamma_{ij}(t)\xi_{ij}(t)
 \left[ \begin{BMAT}{c}{cc} \alpha_{i} y_{i}(t-d) \\ 1 + \beta_{i}y_{i}(t-d) \end{BMAT} \right].$
The following compact form analogous to (\ref{compf}) follows:
\begin{align} \label{rn}
r^{f}(t+1)=&[I+ \delta(t)(\Phi^{f}(t)+\Psi^{f}(t,d)) (\Gamma^{f}(t) \otimes I_{2}) +\delta(t) \tilde{N}^{f}(t,d) ] r^{f}(t) \nonumber \\&+\delta(t)\nu^{f}(t,d)
\end{align}
where $\Phi^{f}(t,d)$, $\Psi^{f}(t,d)$ and $\tilde{N}^{f}(t,d)$ are obtained from $\Phi(t,d)$, $\Psi(t,d)$ and $\tilde{N}(t,d)$ in (\ref{compalgnd}) in the same way as $\Phi^{f}(t)$ is obtained from $\Phi(t)$ in (\ref{compf}),  while $\nu^{f}(t,d)$ $=[\nu_{l_{1}}(t)^{T} \cdots $ $ \nu_{l_{|\mathcal{N}-\mathcal{N}^{f}|}}(t)^{T}]^{T}$. It is important to observe that
$B^{f}(t,d)=\Phi^{f}(t) (\Gamma^{f}(t) \otimes I_{2})=$ $\bar{B}^{f}(d)+\tilde{B}^{f}(t,d)$, where $\bar{B}^{f}(d)=E \{B^{f}(t,d) \}$ which is Hurwitz for some $d=d_{0}$, according to A4'). Also, $\{\Psi^{f}(t,d)\},$ $ \{  \tilde{N}^{f}(t,d)\},$ and $
\{\nu^{f}(t,d)\}$ are random sequences satisfying $E\{ \Psi^{f}(t,d)|$ $ \mathcal{F}_{t-1}\}=0$, $ E \{
\tilde{N}^{f}(t,d)| $ $\mathcal{F}_{t-1} \}=0$ and $E \{\nu^{f}(t,d)| $ $\mathcal{F}_{t-1} \} =0$; however, $E \{
\tilde{B}^{f}(t,d) \}=0$, but $E \{\tilde{B}^{f}(t,d)|\mathcal{F}_{t-1} \} \neq 0$, in general. Using the methodology of Theorem 4 we iterate
(\ref{rn}) $\tau$ steps backwards; then, one can apply the methodology used in the proof of Theorem~8, having in mind A2'). Calculating the Lyapunov function $V(t)= E\{ r(t)^{T} R^{f} r(t) \}$ like in the proof of Theorem~2, where $R^{f}$ is
a positive definite matrix satisfying the Lyapunov equation $R^{f} \bar{B}(d_{0})^{f} +
\bar{B}(d_{0})^{fT}R^{f}=-Q^{f}$, where $Q^{f} > 0$, one can analogously show that $r(t)$ tends to
zero in the mean square sense and w.p.1. Thus the result follows. \hspace*{\fill}\QED
\par

\section{Simulation Results}
In order to illustrate properties of the proposed algorithms, a sensor network with ten nodes has been simulated.
A fixed randomly selected communication graph satisfying A3) has been adopted, and parameters $\alpha_{i}$ and
$\beta_{i}$ have been randomly selected around one and zero, respectively, with variance 0.3.

\begin{figure}
\begin{center}
\includegraphics[width=70mm,keepaspectratio,clip]{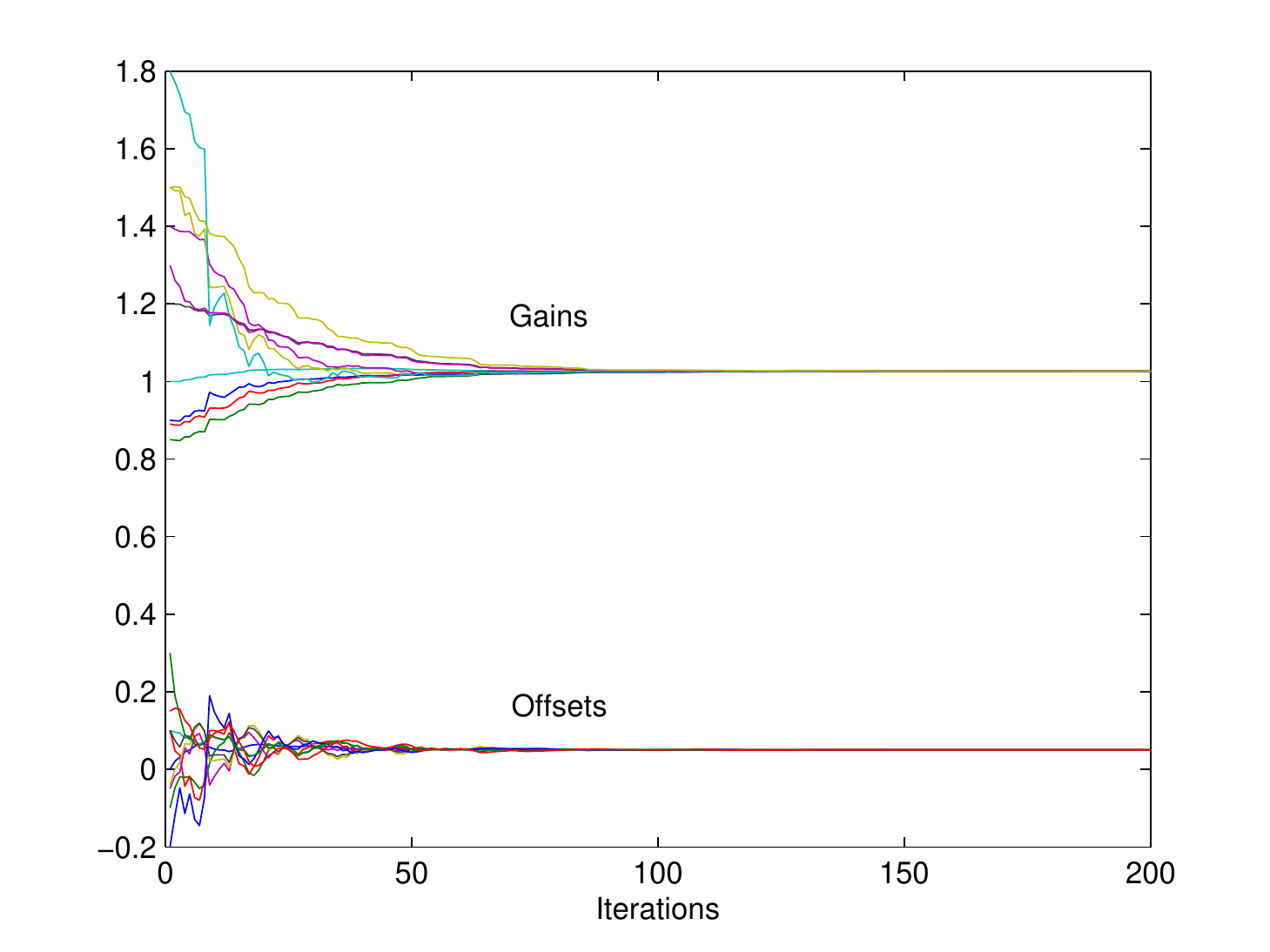}
\end{center}
\caption{Noiseless case: offset and gain estimates with no reference}
\end{figure}

In Fig.~1 the equivalent gains $\hat{g}_{i}(t)$ and offsets $\hat{f}_{i}(t)$ generated by the proposed algorithm \eqref{alg}
are presented for a preselected gain $\delta=0.01$ in the noiseless case. It is clear that calibration is achieved, and that
the asymptotic values are equal; in this case they are close to the optimal values. Fig.~2 depicts the situation when the first node is
assumed to be a reference node with $\alpha_{1}=1$ and $\beta_{1}=0$. Convergence to the reference value is
obvious (see Section 5).

\begin{figure}
\begin{center}
\includegraphics[width=70mm,keepaspectratio,clip]{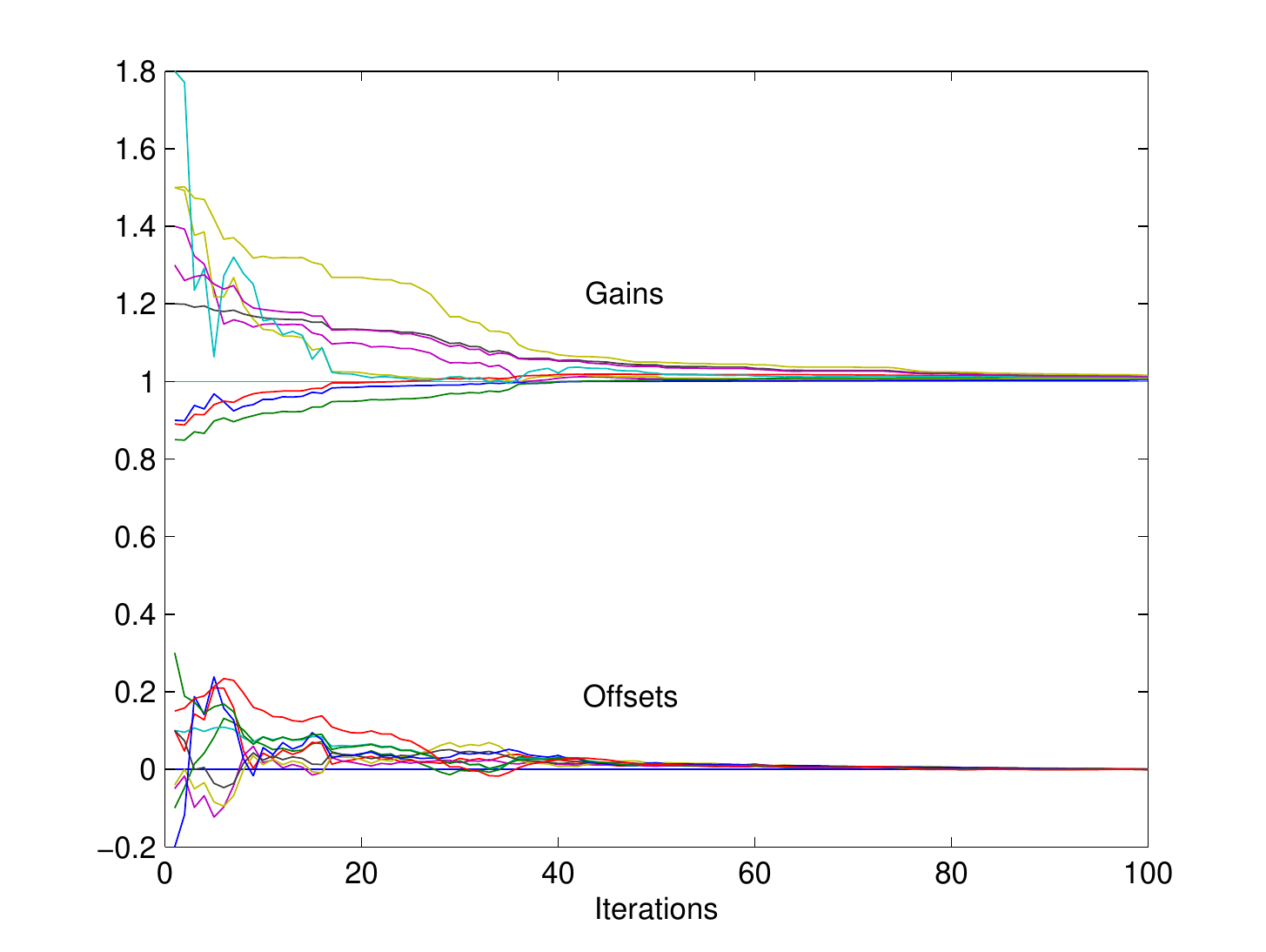}
\end{center}
\caption{Noiseless case: offset and gain estimates with reference included}
\end{figure}

\begin{figure}
\begin{center}
\includegraphics[width=70mm,keepaspectratio,clip]{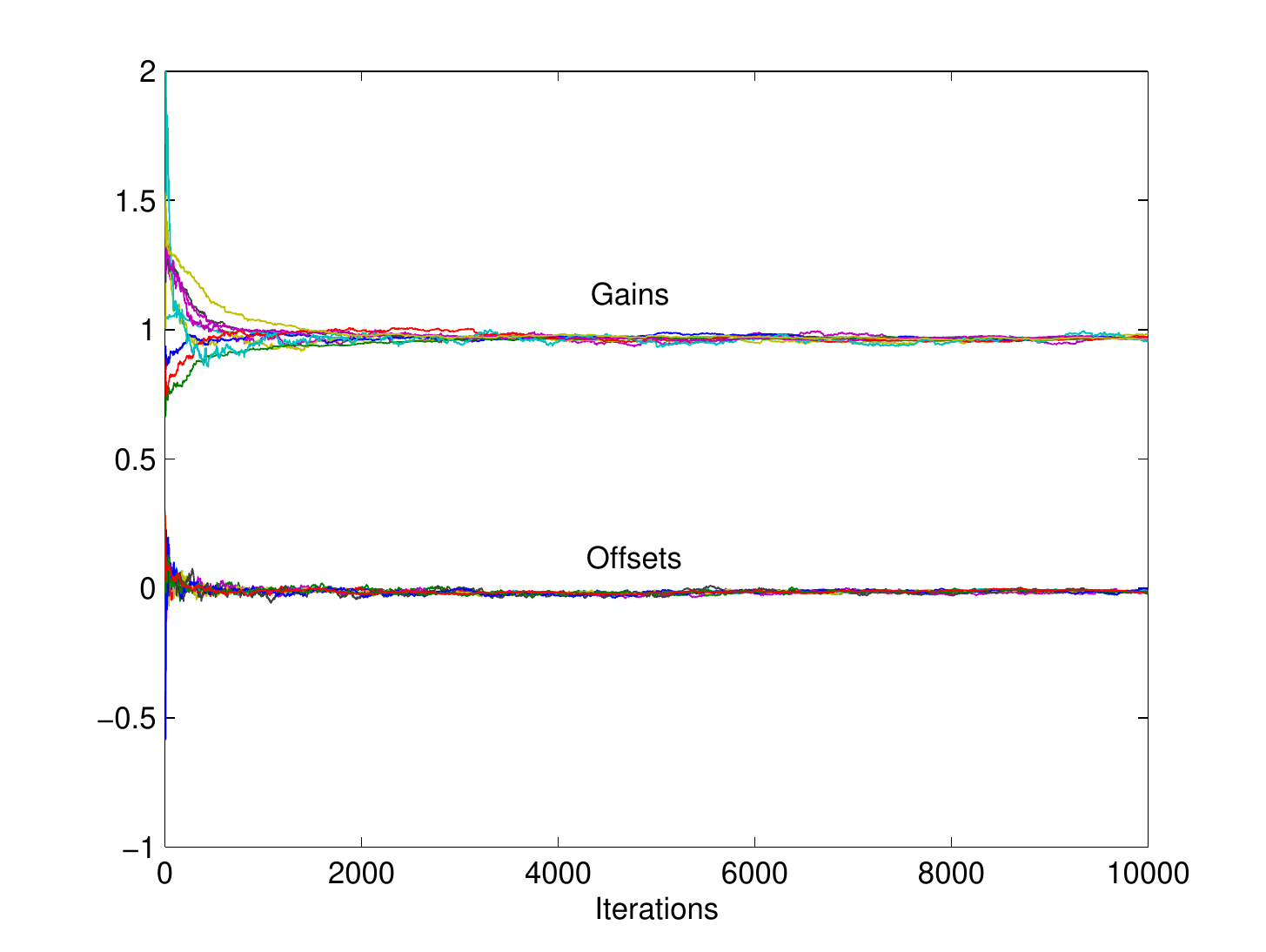}
\end{center}
\caption{Lossy sensor network: offset and gain estimates for instrumental variable method with $d=1$}
\end{figure}

In Fig.~3 the equivalent gains
$\hat{g}_{i}(t)$ and offsets $\hat{f}_{i}(t)$ generated by the proposed instrumental variable algorithm (\ref{alginst}) with $d=1$ are
presented for the sequence $\delta(t)=0.01/t^{0.6}$. All the discussed uncertainties are included:
communication outages with $p=0.2$, communication additive noise with variance 0.1, and measurement noise with
variances randomly chosen in the range $(0,0.1)$; the signal $x(t)$ is a correlated random sequence with zero mean and
variance $1$. It is clear that calibration is achieved in spite of the noise existence.
\begin{figure}
\begin{center}
\includegraphics[width=70mm,keepaspectratio,clip]{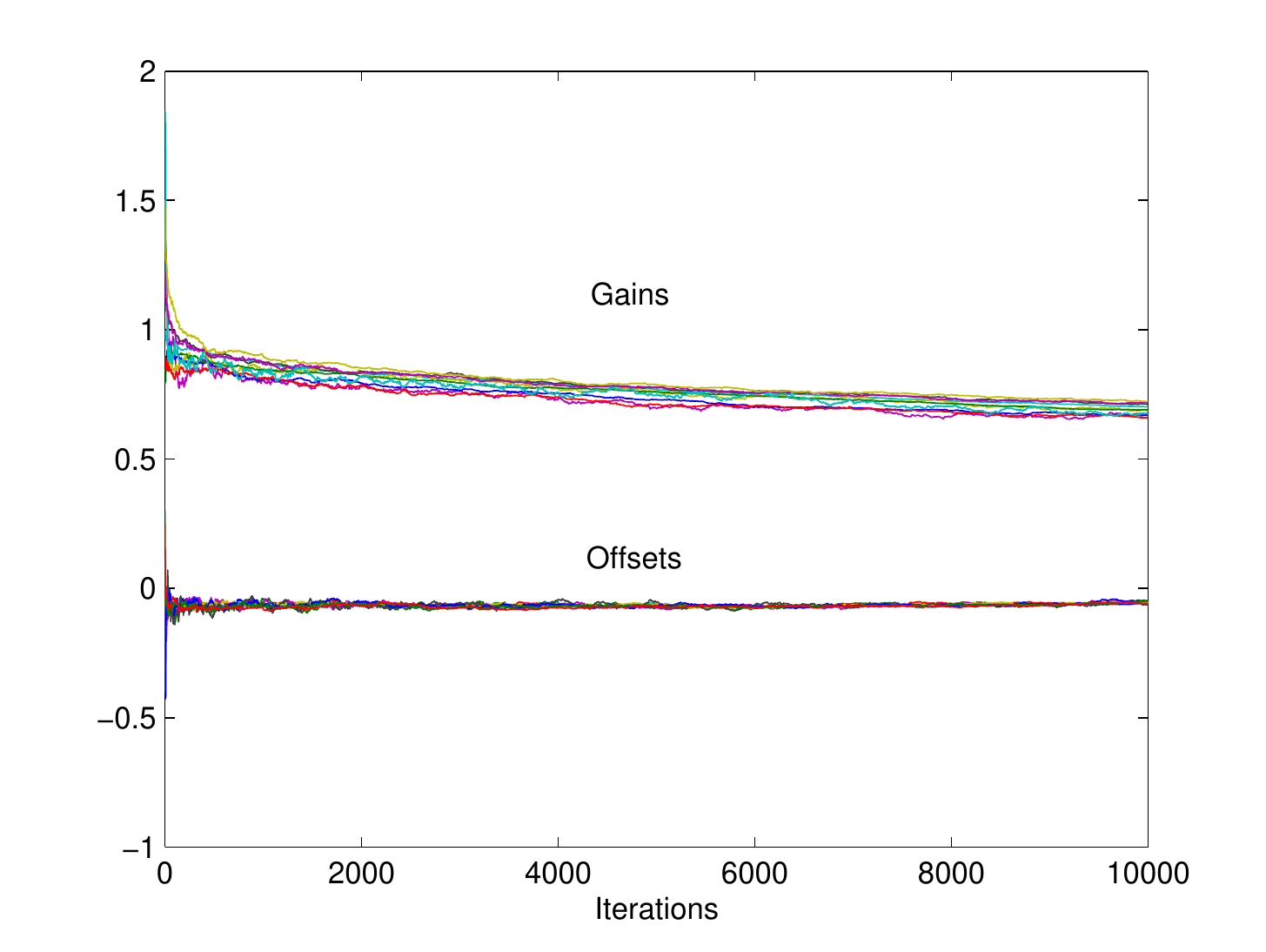}
\end{center}
\caption{Lossy sensor network: offset and gain estimates for $d=0$, \textit{i.e.}, without the proposed modification }
\end{figure}
Fig.~4 illustrates the necessity of introducing instrumental variables in the basic gradient algorithm. Here the estimates are presented in the above noisy case for $d=0$ (no instrumental variables). Obviously, synchronization of the network is not achievable in this case since the equivalent gains $\hat{g}_{i}(t)$ slowly converge to zero.

\section{Conclusion}
In this paper a new \emph{distributed blind calibration algorithm} of gradient type resulting in extended consensus has been proposed for sensor networks. The algorithm provides a new efficient tool for coping with the problem of calibration of large wireless sensor networks with communications limited to close neighbors, without requiring any fusion center. It
is proved, after developing a novel methodology of treating \emph{higher-order consensus schemes} based on diagonal dominance of matrices decomposed into blocks, that the algorithm achieves asymptotic
agreement on all sensor gains and offsets in a given network. Convergence to consensus in the mean square sense and with probability one is proved separately for the cases of noiseless and noisy environments. In the case of imperfect inter-node communications and measurement noise, a modification of the basic gradient algorithm based on the introduction of \emph{instrumental variables} is proposed. It is proved using stochastic approximation arguments that the extended algorithm provides synchronization of both gains and offsets eliminating additive (communication and measurement) noises and in the presence of communication packet dropouts under nonrestrictive assumptions. Special attention is paid to the convergence rate of the proposed algorithms. A special section is devoted to the problem of distributed macro calibration of sensor networks when a subset of nodes have fixed calibration parameters. It is proved that in this case for both noiseless and noisy versions the proposed algorithm ensures convergence to a vector of invariant nodes characteristics. Some simulation results illustrate the behavior of the proposed algorithms.

The presented blind calibration methodology based on gradient recursions derived from (\ref{J}) can also be applied \emph{asynchronously}. For example, one can construct a system in which all the nodes in the network have local clocks that tick independently from each other according to Poisson processes. At each tick of an internal clock the corresponding agent broadcasts its current output to the nodes from the set of its out-neighbors. After receiving the message, each out-neighbor takes its own measurement and makes one iteration of the recursion (\ref{alg}). The procedure continues after the next tick of one of the clocks. Another asynchronous algorithm can be constructed when the network graph is undirected and the communication delays are not significant: at the tick of an internal clock, the agent broadcasts its request to the neighboring nodes. The neighboring nodes make their measurements and send them immediately back to the requesting node, which performs one iteration (\ref{alg}) using its own measurement. All such schemes based on different forms of \emph{gossip} (\textit{e.g.}, \cite{aysal}) have the advantage of not requiring special calibration periods: they can be implemented in parallel with the normal sensor functioning. Convergence of such schemes is out of the scope of this paper. It could be studied following the methodology exposed above (\emph{e.g.}, using the  results in Subsection 3.1). Also, such schemes can eventually allow the extension of the proposed approach to the problem of time synchronization in sensor networks (see \cite{msj_necsys12} for some initial results).

The presented results based on Lemma~2 open up a possibility of extending the obtained results to the case of synchronization of general higher order linear parameter-varying systems. The case when the nodes measure spatially varying signals deserves further attention.

\bibliography{formifaccalib}

\section{Appendix}
\subsection{Proof of Lemma 2}
If $A_{ii}$ is Hurwitz, then there exists a positive definite matrix D, such that $A_{ii}
D+DA_{ii}^{*}=-Q_{D}$, where $Q_{D}$ is positive definite. Define the following operator norm of a matrix $X \in
\mathbb{C}^{m \times m}$
\begin{equation} \label{normd}
\|X \|=\sup_{x \neq 0} \|Xx \|_{D}/\| x \|_{D},
\end{equation}
 where $x \in \mathbb{C}^{m}$, and $\|x
\|_{D}=(x^{*}D^{-1}x)^{\frac{1}{2}}$, while $D > 0$ is such that $Q_{D}
> 0$. Using the norm (\ref{normd}) in the definition of the matrix $W$ in Lemma~\ref{lemma:1}, we have for its off-diagonal elements
\begin{align}
w_{ij}=&-\| A_{ii}^{-1} A_{ij} \| \nonumber \\
=&-\sqrt{\lambda_{\max}(A_{ij}^{*}A_{ii}^{-1*} D^{-1} A_{ii}^{-1}A_{ij})} \nonumber \\
=&-\max_{x \neq 0} \frac{x^{*}A_{ij}A_{ij}^{*}x}{x^{*}A_{ii}DA_{ii}^{*}x}.
\end{align}
According to Lemma~1, $A$ is Hurwitz if $A-\lambda I$ has quasi-dominating diagonal blocks for all $\lambda
\in \mathbb{C}_{+} $, which is satisfied if the following holds
\begin{equation}
x^{*} (A_{ii}- \lambda I)D(A_{ii}- \lambda I)^{*} x \geq x^{*} A_{ii}DA_{ii}^{*} x
\end{equation}
for all $\lambda \in \mathbb{C}_{+} $, since this guarantees that the corresponding matrix $W(\lambda)$ is an M-matrix for all $\lambda$. Let $\lambda=\sigma+j \mu$ be a complex number with a nonnegative real part. Then, we have
\begin{align}
 H &=  x^{*} (A_{ii}- \lambda I)D(A_{ii}- \lambda I)^{*} x \nonumber \\ &\geq x^{*} A_{ii}DA_{ii}^{*} x+ x^{*} (A_{ii}Dj -DA_{ii}^{*} j)x \mu \nonumber \\ & \quad
- x^{*} \sigma (A_{ii}D+DA_{ii}^{*}) x +  \mu ^{2} \lambda_{\min}(D) x^{*}x.
\end{align}
As $x^{*} (A_{ii}D - DA_{ii}^{*} )x = 0$ and $x^{*} \sigma (A_{ii}D+DA_{ii}^{*})x \leq 0$ for $\sigma \geq 0$
according to the assumption of the Lemma, we have that $H \geq x^{*} A_{ii}DA_{ii}^{*} x$. Hence, the result follows.

\subsection{Proof of Lemma 3}
Let $W^{d}=[w^{d}_{ij}]$, where $w^{d}_{ij} = 0$ for $i=j$, and $w^{d}_{ij}=
(\sum_{j=1,j \neq i}^{n} \gamma_{ij})^{-1}\gamma_{ij}$ for $i \neq j$. This matrix is row stochastic and
cogredient (amenable by permutation transformations) to
\begin{equation} \label{wc}
W^{d}_{c}=\left[ \begin{BMAT}{cc}{cc} W^{d}_{1} & 0 \\ W^{d}_{2} & W^{d}_{0} \end{BMAT}
\right],
\end{equation}
where $W^{d}_{1} \in \mathbb{R}^{n_{1} \times n_{1}}$ is an irreducible matrix, $W^{d}_{2} \in \mathbb{R}^{n_{2} \times
n_{1}} \neq 0$ and $W^{d}_{0} \in \mathbb{R}^{n_{2} \times n_{2}}$ is such that $\max_{i}$ $ |\lambda_{i} \{
W^{d}_{0} \}| < 1$. Select one node of the graph $\mathcal{G}$ from the set of center nodes and delete the corresponding row and
column from $\Gamma$; the resulting matrix is cogredient to $
W^{d-}_{c}=\left[ \begin{BMAT}{cc}{cc} W^{d-}_{1} & 0 \\ W^{d-}_{2} & W^{d}_{0} \end{BMAT}
\right],$ where $W^{d-}_{1} \in \mathbb{R}^{(n_{1}-1) \times (n_{1}-1)}$ and $W^{d-}_{2} \in \mathbb{R}^{n_{2} \times
(n_{1}-1)}$. As $W^{d}_{1}$
corresponds to a closed strong component of the inverse graph of $\mathcal{G}$, deleting one node from it
(together with the corresponding edges) results in a graph containing, in general, $\kappa \geq 1$ closed strong
components; each of the nonnegative matrices to which these strong components are associated have at least one row with the sum of all the elements strictly less
than one. Consequently, $I-W^{d -}_{c}$ is an M-matrix \cite{s}. Deleting one row and column from $\Gamma$ means deleting two consecutive rows and columns from  $\bar{B}$; let $\bar{B}^{-}\in \mathbb{R}^{(2n-2) \times (2n-2)}$ be the resulting matrix. According to Lemmas 1 and 2 and assumption A4) which guarantees that  $\bar{\Phi}_{i}$, $i=1, \ldots, n$ is Hurwitz,  $\bar{B}^{-}$ has quasi-dominating
diagonal blocks, which directly implies that $\bar{B}^{-}$ has all the eigenvalues with negative real parts. Thus the result.

\subsection{Proof of Lemma 4}
The eigenvalue of $\bar{B}$ at the origin has both algebraic and geometric multiplicity equal to
two: $i_{1}$ and $i_{2}$ represent two corresponding linearly independent eigenvectors. The rest of the proof
follows from the Jordan decomposition of $\bar{B}$.  Matrix $\bar{B}^{*}$ is Hurwitz
according to Lemma~\ref{lemma:3}.
\subsection{Proof of Lemma 5}
We conclude immediately that vectors $i_{1}$ and $i_{2}$ are eigenvectors for both
$\bar{B}$ and $B(t)$, taking into account (\ref{compalgm}) and (\ref{compalg}).

Let $w=\left[
\begin{BMAT}{ccc}{c} w_{1} \cdots w_{2n}
\end{BMAT} \right]$ be a left eigenvector of $\bar{B}$ corresponding to the zero eigenvalue (either $\pi_1$ or $\pi_2$). Then, $w \bar{B}=0$ gives :
\begin{align} \label{w1}
0=&-[w_{2i-1}(\alpha_{i} \beta_{i}+\alpha_{i}^{2} \bar{x})+ w_{2i}(1+\beta_{i}^{2}+\alpha_{i} \beta_{i} \bar{x})] \sum_{j=1, j \neq i}^{n} \gamma_{ij} \nonumber \\ & +  \sum_{l=1, l \neq i}^{n}
[w_{2l-1}(\alpha_{l} \beta_{l}+\beta_{l}^{2} \bar{x}) + w_{2l}(1+\beta_{l}^{2}+\alpha_{l}
\beta_{l} \bar{x})] \gamma_{li},
\end{align}
\begin{align} \label{w2}
0=& -[w_{2i-1}(\alpha_{i} \beta_{i} \bar{x}+\alpha_{i}^{2} s^{2})+w_{2i}((1+\beta_{i}^{2})\bar{x}+\alpha_{i}
\beta_{i} s^{2})]  \sum_{j=1, j \neq i}^{n} \gamma_{ij} \nonumber \\ &+  \sum_{l=1, l \neq i}^{n}
[w_{2l-1}(\alpha_{l} \beta_{l} \bar{x}+\beta_{l}^{2} s^{2}) + w_{2l}((1+\beta_{l}^{2})\bar{x}+\alpha_{l} \beta_{l} s^{2})] \gamma_{li},
\end{align}
for $i=1, \ldots, n$. We simply conclude from (\ref{w1}) and (\ref{w2})  $w \bar{B} =0 \Longrightarrow w B(t)=0$,
since the components of $v(t)=w B(t)$ are
\begin{align} \label{w1b}
v_{2i-1}(t)=& -[w_{2i-1}(\alpha_{i} \beta_{i}+\alpha_{i}^{2}x(t))+w_{2i}(1+\beta_{i}^{2}+ \alpha_{i} \beta_{i}
x(t))] \sum_{j=1, j \neq i}^{n} \gamma_{ij} \nonumber \\ & +  \sum_{l=1, l \neq i}^{n} [w_{2l-1}(\alpha_{l} \beta_{l}+\beta_{l}^{2} x(t)) +  w_{2l}(1+\beta_{l}^{2}+\alpha_{l} \beta_{l} x(t))] \gamma_{li} = 0,&
\end{align}
\begin{align} \label{w2b}
v_{2i}(t)=& -[w_{2i-1}(\alpha_{i} \beta_{i}x(t)+\alpha_{i}^{2}x(t)^{2})\nonumber \\ &+w_{2i}((1+\beta_{i}^{2})  x(t)+
\alpha_{i} \beta_{i} x(t)^{2})]\sum_{j=1, j \neq i}^{n} \gamma_{ij}  \nonumber \\
&+ \sum_{l=1, l \neq i}^{n} [w_{2l-1}(\alpha_{l} \beta_{l}x(t)+\beta_{l}^{2} x(t)^{2}) \nonumber \\ & + w_{2l}((1+\beta_{l}^{2})x(t)+\alpha_{l} \beta_{l} x(t)^{2})] \gamma_{li}  =0,
\end{align}
$i=1, \ldots, n$, having in mind that (\ref{w1}) and (\ref{w2}) imply \[-(w_{2i-1} \alpha_{i}^{2}+ w_{2i} \alpha_{i} \beta_{i} ) \sum_{j=1, j \neq i}^{n} \gamma_{ij} + \sum_{l=1, l \neq i}^{n}
(w_{2l-1} \alpha_{l}^{2}  + w_{2l} \alpha_{l} \beta_{l}) \gamma_{li}=0\]
and \[-[w_{2i-1} \alpha_{i} \beta_{i}+ w_{2i}(1+\beta_{i}^{2})] \sum_{j=1, j \neq i}^{n} \gamma_{ij}+ \sum_{l=1, l \neq i}^{n}[w_{2l-1} \alpha_{l} \beta_{l}+w_{2l}(1+\beta_{l}^{2})]\gamma_{li}=0\]
by virtue of A4).  Therefore, we have $\pi_{1}B(t)=0$ and $\pi_{2} B(t)=0$, and the result follows taking into
account (\ref{trans}).
\end{document}